\documentclass[preprint,12pt]{elsarticle}

\usepackage{amssymb}
\usepackage{amsmath}
\usepackage[figuresright]{rotating}
\usepackage{bm}
\usepackage{caption}
\usepackage{multicol}
\usepackage{algorithm}
\usepackage{algpseudocode}
\usepackage{fancyhdr}
\usepackage{geometry}
\usepackage{url} 
\usepackage[figuresright]{rotating}
\usepackage{bm}
\usepackage{array}
\usepackage{caption}
 \usepackage{booktabs}
\usepackage{titlesec}
\titleformat*{\section}{\small \bf}
\titleformat*{\subsection}{\small \em}
\titleformat*{\subsubsection}{\small \em}

\journal{Journal}

\begin{document}

\begin{frontmatter}



\title{
\begin{flushleft}
{\LARGE Large Language Model Based Multi-Agent System Augmented Complex Event Processing Pipeline for Internet of Multimedia Things}
\end{flushleft}
}



\author[]{ \leftline {Talha Zeeshan$^a$, Abhishek Kumar$^*$$^a$, Lauri Lovén$^a$, Susanna Pirttikangas$^a$, Sasu Tarkoma$^b$$^a$}}

\address{ \leftline {$^a$Center for Ubiquitous Computing, University of Oulu, 90570 Oulu, Finland}

  \leftline {$^b$Department of Computer Science, University of Helsinki, 00560 Helsinki, Finland}

}

\cortext[]{Abhishek Kumar (Corresponding author). Address: Center for Ubiquitous Computing, Erkki Koiso-Kanttilan katu 3, P.O Box 4500, University of Oulu, Linnanmaa, 90570 Oulu, Finland. Email: \url{abhishek.kumar@oulu.fi}.}


\begin{abstract}
This paper presents the development and evaluation of a Large Language Model (LLM), also known as foundation models, based multi-agent system framework for complex event processing (CEP) with a focus on video query processing use cases. The primary goal is to create a proof-of-concept (POC) that integrates state-of-the-art LLM orchestration frameworks with publish/subscribe (pub/sub) tools to address the integration of LLMs with current CEP systems. Utilizing the Autogen framework in conjunction with Kafka message brokers, the system demonstrates an autonomous CEP pipeline capable of handling complex workflows. Extensive experiments evaluate the system's performance across varying configurations, complexities, and video resolutions, revealing the trade-offs between functionality and latency. The results show that while higher agent count and video complexities increase latency, the system maintains high consistency in narrative coherence. This research builds upon and contributes to, existing novel approaches to distributed AI systems, offering detailed insights into integrating such systems into existing infrastructures. 

\end{abstract}


\end{frontmatter}


\section{Introduction}
The rapid advancement of artificial intelligence (AI) technologies has revolutionized the way we process and analyze data, particularly in the field of complex event processing, such as video query analysis. Traditional CEP systems often struggle with the dynamic demands of modern applications such as real-time or near real-time video analytics that require the integration of diverse data sources, for example, thousands of surveillance cameras deployed in a city, leading to limitations in their performance and applicability. 

Modern CEP pipelines are domain-specific and often struggle to adapt to dynamic changes in the environment in a timely manner. State-of-the-art applications (such as live video streaming on TikTok, YouTube etc.) generate an increasing volume of diverse, complex data that needs to be handled in the appropriate manner depending on the use case. Large Language Models (LLMs), also known as foundation models, inherently possess the ability to handle and analyze dynamic forms of data and therefore provide the necessary foundation upon which a dynamic CEP pipeline can be created which can support a diverse range of domains. Our work is derived from this growing need for automated efficient and scalable complex event processing (CEP) systems capable of handling dynamic changes and adapting to changing scenarios. Hence, the primary motivation behind our work is to introduce the usage of LLM agents to augment existing CEP pipelines and explore the potential of LLM-based multi-agent systems to enhance the capabilities of CEP frameworks. By leveraging the strengths of LLMs in such a system, we aim to create a more robust and flexible CEP framework that can fulfil these changing requirements.

This paper focuses on two key research questions: 

\textit{RQ1: What is the current landscape of LLM-based single or multi-agent systems?} LLM-based agent framework is a rapidly emerging field of research. The objective is to provide a comprehensive literature review that highlights the various applications and domains of LLM agent systems.

\textit{RQ2: How can Large Language Models (LLMs) be integrated with multi-agent systems to augment complex event processing (CEP) pipelines that meet stringent quality of service (QoS) metrics associated with the given use case?} Integrating LLM-MAS with existing pub/sub infrastructures such as Kafka, necessitates a strategic approach to ensure seamless transition and operation. Given that LLM-backed agents introduce a novel approach to automate existing implementations, it is important to ensure that the proof-of-concept (POC) interfaces effectively with current systems, tools, and technologies. This ensures compliance with existing standards and minimizes disruption to established workflows.

With these research questions in mind, we study the feasibility of utilizing an LLM-MAS framework such as Autogen in tandem with Pub/Sub tools such as Kafka to satisfy the requirement and present a proof-of-concept (POC) that unifies the state-of-the-art LLM Multi-Agent system (LLM-MAS) framework and Pub/Sub tools and technologies to create a CEP pipeline. Our contributions are as follows:

\begin{enumerate}
    \item A comprehensive survey, gap analysis of LLM-based autonomous agent applications.
    \item A proof-of-concept for an LLM-MAS augmented CEP pipeline with pub/sub broker tools that can be tested, evaluated and expanded.
    \item A discussion of lessons learned from deploying multi-agent-based systems for CEP use cases.

\end{enumerate}

The rest of this paper is organized as follows: section~\ref{sec:background} describes the background technologies relevant to the focus of this paper; section~\ref{sec:related work} provides a comprehensive survey, gap analysis and future trend of LLM-based autonomous agent applications; section~\ref{sec:system-design} presents system design and implementation of our proof-of-concept; section~\ref{sec:exp} presents experimental protocols, use case, results along with the discussion and limitations; and finally section~\ref{sec:conclusion} provides concluding remarks along with future research directions.







\section{Background}\label{sec:background}
This section presents background  of key technologies relevant to the focus of our work.
\subsection{Distributed AI Systems and the Computing Continuum}
The current state-of-the-art in the Distributed AI (DAI) and computing continuum is primarily characterized by the trend of how AI systems can be designed and deployed while utilizing horizontal resource scaling offered by the continuum. These trends focus on a variety of domains that include the integration of AI with edge computing (aka. Edge AI), advancements in federated learning, the rise of TinyML (AI/ML for IoT), and the adoption of blockchain technology for enhanced security and decentralization in AI systems~\cite{kokkonen2022autonomy}. 

The computing continuum can be described as mixture of cloud, edge, fog computing and IoT that allows for real-time data processing and decision making at the edge of the network~\cite{Zou2019}. The concept of the computing continuum represents an advanced model in the evolution of distributed computing frameworks. It extends beyond the traditional cloud-centric models by incorporating a wide array of computational resources that span from centralized cloud data centers, towards decentralized, multi-tier infrastructures at the edge of the network~\cite{rosendo2022}, and even to endpoint devices like smartphones and IoT devices. This seamless integration of computational resources across a spectrum allows for more efficient data processing, lower latency, and enhanced scalability~\cite{buyya2018manifesto}. The computing continuum, encompassing a range of computing environments from the cloud to the edge, has significantly transformed the capabilities of AI applications. By distributing computational resources across this continuum, AI applications can now leverage the right kind of computing power at the right location, leading to increased efficiency, reduced latency, and enhanced scalability. Applications that require immediate responses, such as autonomous vehicles, emergency response systems, and real-time financial trading algorithms, benefit immensely from reduced latency. By processing data closer to where it is generated (at the edge), these applications can react to real-world events in a timely manner, a feat that would be challenging with cloud-only models~\cite{satyanarayanan2017}. 

\subsection{Publish/Subscribe Paradigm}
\begin{figure}[t]
  \centering
   \includegraphics[width=0.8\linewidth]{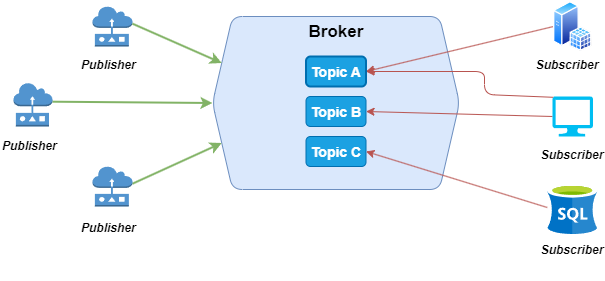}
   \caption{A General Publish/Subscribe Architecture}
   \label{fig:pub-sub}
\end{figure}

The Publish/Subscribe (Pub/Sub) paradigm is a messaging model predominantly utilized in distributed systems~\cite{eugster2003, tarkoma2012publish}. It is characterized by a decoupled interaction model where the entities producing the messages (publishers) are separated from the entities receiving these messages (subscribers)~\cite{eugster2003}. This model is instrumental in systems where the dissemination of information is event-driven and requires broad yet targeted distribution~\cite{Carzaniga2001}. Figure~\ref{fig:pub-sub} showcases the general architecture of a publish/subscribe model.

The core concepts encompassing the Pub/Sub paradigm can be summarized as follows:
\begin{itemize}
    \item \textbf{\textit{Publishers:}} Publishers are responsible for generating and sending out messages. Crucially, they operate independently of the subscribers, often unaware of the recipients of the information they publish.
\end{itemize}

\begin{itemize}
    \item \textbf{\textit{Subscribers:}} Subscribers express their interest in certain types of messages by subscribing to specific topics or event channels. They receive messages that match their subscription criteria, enabling them to process relevant information as needed.
\end{itemize}

\begin{itemize}
    \item \textbf{\textit{Event Channels/Topics:}} These are the pathways or categories to which messages are published. Publishers send messages to these channels, and subscribers receive messages from the channels they have subscribed to.
\end{itemize}

\begin{itemize}
    \item \textbf{\textit{Event Broker/Message Broker:}} This central component, often a server or a set of servers, manages the distribution of messages. It filters and routes the messages from publishers to the appropriate subscribers, ensuring that each subscriber receives only the content of interest \cite{Team_2015}.
\end{itemize}

The Pub/Sub model provides key features that contribute to its widespread adoption and effectiveness. One of the fundamental advantages of the Pub/Sub paradigm is the decoupling of publishers and subscribers~\cite{eugster2003}, which separates the roles of message producers (publishers) and consumers (subscribers). This separation allows for modifications or extensions to either publishers or subscribers without necessitating changes to the other, hence fostering a flexible and adaptable architecture. Another pivotal characteristic is its support for asynchronous communication, where publishers and subscribers do not need to interact with the system simultaneously~\cite{saleh2023pubsub}. This aspect is particularly advantageous in distributed environments where components may operate under varied schedules or in different time zones, ensuring continuous and uninterrupted information flow. Furthermore, the Pub/Sub paradigm is marked by its robust filtering mechanisms, which empower subscribers to specify and receive only the messages that align with their interests or criteria~\cite{eugster2003, saleh2023pubsub}. Such targeted delivery is instrumental in maintaining the efficiency and relevance of the communication process, avoiding the inundation of subscribers with extraneous data.



\subsection{Large Language Models (LLMs) / Foundation Models}
\label{sec:llms}
Large Language Models (LLMs) have brought upon a revolution in terms of how we think of and approach artificial intelligence. These large models can take in images, text and audios as inputs and produce textual outputs~\cite{openaigpt42024, Touvron2023LLaMAOA}, providing powerful tools for natural language processing, reasoning, and decision-making tasks~\cite{emergentAbilities}.

LLMs can be described as neural network models that are trained on enormous volume of text corpus and can have millions or billions of parameters. For context, models like BERT (Bi-directional Encoder Representation from Transformers)~\cite{devlin2019bert} is composed of 110 million parameters, ChatGPT which was based on the GPT-3.5 model at the time of its public release, boasted 175 billion parameters~\cite{bahrini2023chatgpt}, LLaMA~\cite{Touvron2023LLaMAOA} is a collection of LLMs developed by meta that boast models whose parameters range from 7 billion to 70 billion and PaLM-2~\cite{anil2023palm} possesses up to 340 billion parameters.

Most modern LLMs are based on the Transformer architecture that provides unmatched capacity and parallelization~\cite{zheng2023understanding}. The transformer architecture was first introduced by Vaswani et.al~\cite{vaswani2023attention} that relies exclusively on self-attention mechanisms and was the foundation of first generation LLMs. However, most current high-performing LLMs (listed above) are based on the casual-decoder architecture. This architecture can be described as a decoder-only design that utilizes a diagonal mask matrix so that the current token can only access information from tokens earlier in the stack~\cite{zheng2023understanding}. This break through in LLM architecture and design led to the creation of high-performing LLMs such as GPT-4 which proved to be a disruptive technology. With the advent of such models, new research avenues were primed for exploration and use-cases across all-domains experienced disruptive effects due to the power of LLMs. Liu et.al~\cite{Liu_2023_gpt} in their paper highlight the affect of Chat-GPT on various research domains, ranging from audio-speech processing, economics, astrophysics and cryptography.  Huang et al.~\cite{huang2022language} in their paper, showcased the potential of LLMs to function as zero-shot planners, extracting actionable knowledge for agents in an interactive environment. Their work focused on grounding high-level tasks, expressed in natural language, into a series of actionable planned steps. They found that large pre-trained language models can effectively decompose high-level tasks into actionable plans. For example, in their environment, a high level tasks such "make fried eggs" can be decomposed into an actionable plan such as; "Open Fridge, Take out 2 eggs from the egg carton, Place pan on stove, add oil, ...". Continuous advancement in LLMs lead to the creation of multi-modal LLMs that can process not just textual input, but other forms of media such as audio-video, thus opening the door to research based in complex-event processing.

\subsection{Autogen: An LLM Multi-Agent Framework}
\label{sec:autogen}
AutoGen is an open-source framework that enables the development of applications through multi-agent conversations~\cite{wu2023autogen}. This section explores the AutoGen LLM orchestration framework, detailing its architecture, functionality, and usage in the presented proof-of-concept.

Autogen allows for the creation of LLM applications by utilzing "Agents" that are capable of inter-agent communication. Agents can be described as entities that have LLM capabilities and are able to send and receive messages from other agents. Each agent can be configured to execute specific tasks within a workflow, such as code generation, code execution, or syntax validation~\cite{wu2023autogen}. This inter-agent communication capability enables the creation of complex workflows that surpass the capabilities of single-agent frameworks. AutoGen facilitates the creation of LLM applications that utilize multiple agents capable of conversing with each other to accomplish a diverse range of tasks. The framework supports the integration of LLMs, human inputs, and external tools and technologies to create agents that can operate in a flexible, modular, and customizable manner~\cite{wu2023autogen}. 

By leveraging recent advances in chat-optimized LLMs like GPT-4~\cite{openaigpt42024}, Llama~\cite{Touvron2023LLaMAOA}, AutoGen allows these agents to cooperate, reason, and validate each other's outputs through structured conversations. This multi-agent orchestration framework thus provides a robust foundation for developing sophisticated LLM applications, where agents can dynamically interact and cooperate to achieve complex objectives.

\subsubsection{Autogen Agents}
\label{sec:autogen_agents}
AutoGen's flexible and modular architecture is built around the concept of "conversable agents," which are designed to handle specific roles within a multi-agent conversation. These agents can send and receive messages, maintain context, and execute tasks based on their configured capabilities. Augogen has three foundation agents each with its own special role and purpose listed below:

\begin{itemize}
    \item \textbf{Conversable Agent:} Serves as the base class for other agents. It can provide automated replies based on feedback and/or code executions. It can generate and execute code and converse with other agents.
    
    \item \textbf{Assistant Agent:} A representative subclass of the conversable agent which is designed to act as an AI assistant. It possesses LLM capabilities and does not require human input. It can receive code execution results and provide feedback for further steps which can guide a complex workflow.
    
    \item \textbf{User Proxy Agent:} Acts as a proxy for the user a multi-agent system. It can prompt for human inputs and executes code/function calls automatically based on human or tool feedback.
    
\end{itemize}
Based on the aforementioned types of agents in Autogen, the kind of agents that can be configured in AutoGen can be broadly categorized into \textbf{LLM-backed agents}, \textbf{human-backed agents}, and \textbf{tool-backed agents}. Each type of agent has unique functionalities and can be customized to meet specific application needs. Each type of agent is further described below:
\newline
\textbf{LLM Backed Agents:}
LLM-backed agents leverage the capabilities of advanced large language models to perform a variety of functions. These agents can role-play, infer implicit states, provide feedback, adapt based on feedback, and even generate and execute code. The primary advantage of LLM-backed agents is their ability to harness the sophisticated language processing and reasoning capabilities of models like GPT-4. They can be configured to handle complex conversational tasks autonomously. Key capabilities of LLM-backed agents include:
\begin{itemize}
    \item \textbf{Role Playing:} These agents can assume specific roles in a conversation, such as a tutor, assistant, or expert, and provide responses based on their role.
    
    \item \textbf{State Inference:} They can infer the state of a conversation from the context due to their shared memory and make decisions accordingly.
    
    \item \textbf{Feedback Provision and Adaptation:} LLM-backed agents can provide feedback on their actions and adapt their responses based on the feedback they receive.
    
    \item \textbf{Code Generation and Execution:} These agents can generate code to solve problems, execute the code, and interpret the results.
\end{itemize}

\textbf{Human Backed Agents:}
Human-backed agents are designed to incorporate human input into the conversation at various stages. This type of agent is crucial for tasks that require human judgment, expertise, or oversight. Human-backed agents can prompt for inputs from human users, ensuring that the conversation or task execution benefits from a human involvement. Configurations and use cases for human-backed agents include:
\begin{itemize}
    \item \textbf{User Proxy:} Acts as a proxy for human users, allowing human input to be incorporated at specific points in the conversation.
    \item \textbf{Human Oversight:} These agents can request human oversight to validate outputs and make decisions, or handle exceptions that are beyond the capabilities of LLMs if the tasks are related to coding.
\end{itemize}

\textbf{Tool Backed Agents:}
Tool-backed agents extend the functionality of AutoGen by enabling the use of external tools and code execution within the conversation. These agents can execute predefined functions, interact with software tools, and manage data processing tasks. By leveraging existing code bases and tools, tool-backed agents enhance the robustness and reliability of the system while minimizing the risks associated with dynamic code generation. Capabilities of tool-backed agents include:
\begin{itemize}
    \item \textbf{Code Execution:} They can execute code snippets generated by LLM-backed agents or provided by human users.
    \item \textbf{Function Calls:} These agents can make function calls to external tools or services, integrating their outputs into the conversation.
    \item \textbf{Error Handling and Debugging:} Tool-backed agents can handle errors during code execution and provide debugging information to resolve issues
\end{itemize}

The integration of these diverse agents within AutoGen allows for the creation of sophisticated multi-agent systems capable of handling a wide range of tasks and applications. By combining different types of agents, AutoGen can support complex workflows that require diverse capabilities. For instance, an LLM-backed assistant agent can generate a solution, which is then validated by a human-backed user proxy agent or executed by a tool-backed agent. This modular approach ensures that each part of the workflow is handled by the most appropriate agent, enhancing the overall efficiency and effectiveness of the system.

\section{Related Works}\label{sec:related work}
This section presents a comprehensive review of prior research and literature in the field of LLM-based autonomous agent applications. By examining the latest advancements in this rapidly evolving domain, this section aims to shed light on the state-of-the-art methodologies and challenges. The objective is to provide a clearer understanding of the current landscape and potential future directions in LLM-based autonomous agents.
\subsection{LLM Agents}
As elaborated upon in Section~\ref{sec:llms}, LLMs have revolutionized the way we interact with artificial intelligence and have proven to be a disruptive technology. They exhibit or show potential in having near human-like intelligence and are the closest thing to Artificial General Intelligence (AGI). With this surge in LLM capabilities, interest has risen in exploring and utilizing LLMs in complex work-flows to achieve a diverse range of tasks. 
An LLM based agent can be considered to possess three main conceptual blocks; \emph{Brain}, \emph{Perception} and \emph{Action}~\cite{appraising2024}. The perception block can be considered to be the "Input" for the agent. The "Brain" represents the LLM model that processes the input and provides an output that can be considered an "Action". 

The development of LLM agents has opened new research avenues across various domains. Given their human-like conversational patterns, there now exists the potential to automate complex processes through simple textual prompts from end users. These LLM agents enable the abstraction of complexities inherent in deconstructing complex events into simpler components and constructing pathways for event resolution. By leveraging the power of LLM agents, it is possible to automate these processes and allow the models to infer the optimal pathways for event resolution. 

LLM agents have captured the attention of researchers across various domains, resulting in a rich body of academic literature. To understand the diverse workflows and complex event processing facilitated by LLM agents, we will divide the exploration and study of these agents into two broad categories: Single-Agent and Multi-Agent systems

\subsection{Single Agent LLM Systems}
OpenAI's GPT model marked a significant breakthrough in the capabilities of large language models (LLMs). By providing access to their API, OpenAI enabled the development of customized plugins to create domain-specific ChatGPT agents~\cite{openaigpt42024}. This innovation facilitated the emergence of specialized Single-Agent Systems (SASs). We define Single-Agent Systems as those that operate independently and lack the capability for multi-agent collaboration.
Due to their singular nature, SASs are inherently limited in the complexity of workflows they can manage. The constraints of operating as a single entity restrict these systems from handling more intricate tasks that require coordination and collaboration among multiple agents. However, they are adept at handling domain specific tasks such as planning, assisting, writing etc. Here, we present existing research that present, highlight and evaluate single agent systems.

Auto-GPT\footnotemark\footnotetext{https://docs.agpt.co/} is an open-source autonomous agent that leverages LLMs to achieve decision-making or planning tasks for real-world scenarios. It follows a single agent paradigm and allows for the creation of novel tools that can help with a given task. Dai et al.~\cite{dai2023adautogpt} introduces and explores the AD-AutoGPT agent, that adapts the Auto-GPT architecture for Alzheimer's Disease (AD) infodemiology. AD-AutoGPT autonomously collects, processes, and analyzes health-related data about AD, leveraging advanced capabilities of GPT-4 to streamline the complex pipeline traditionally used in public health research. The model automatically generates a robust data collection and processing workflow via user textual prompts, significantly reducing the manual effort required and enhancing the comprehensiveness of the data analysis. Similarly, Yang et al.~\cite{yang2023autogpt} also utilizes the Auto-GPT architecture to introduce an agent for online decision making tasks. The proposed agent uses state-of-the-art LLMs (such as GPT-4, GPT-3.5) for realistic scenarios such as online shopping.

Lang-chain\footnotemark\footnotetext{https://python.langchain.com/v0.2/docs/introduction/} is another general framework, similar to Auto-GPT, that allows for the creation of LLM based applications. Lang-Chain is a popular framework for creating LLM applications due to its robust offerings and accessibility. Topsakal et al.~\cite{langChain} provide an in-depth exploration of lang-chain and how its modular abstractions and customizable pipelines enable the development of LLM-based applications. The paper showcases three examples in which lang-chain can be used to develop a single agent system; Information Extraction, Question/Answering, Customized Goal Oriented Agent~\cite{langChain}.

\subsection{Multi-Agent LLM Systems}
As the increasing capabilities of LLMs broadened research horizons, the limitations of single LLM-based agents soon became apparent. LLM agents, by themselves, are limited in what actions they can take since they can only output text. The concept of multi-agents was introduced to harness the collective capabilities of multiple LLM-based agents working in tandem to overcome the constraints of single-agent systems. By distributing tasks among various specialized agents, multi-agent systems can perform more complex and dynamic operations that single agents cannot achieve alone. Here we present some existing applications that utilize multi-agent strategy and highlight their capabilities. Due to the expansive range of work-flows that can now be automated due to LLM-MASs, we divide these systems into two different subsections to provide clarity.

\subsubsection{Planning Workflows}
Further exploring the roles and capabilities of MASs, Paper~\cite{nascimento2023} introduces a novel paradigm that integrates LLMs (such as GPT-4) into these MASs to enhance agent communication and autonomy. Their approach is an extension of the MAPE-K (Monitoring, Analyzing, Planning, Executing, and Knowledge) protocol introduced by IBM in 2004~\cite{nascimento2023}. Nascimento et al. grounded their approach in the MAPE-K model due to its robustness in system adaptability within dynamic environments. This methodology is validated through a practical application in a marketplace scenario, showcasing improved communicative capabilities and adaptability of the agents. This work represents a significant advancement in MAS self-adaptation, indicating further research opportunities to assess LLMs’ applicability in more complex scenarios, potentially leading to enhanced problem-solving capabilities and refined communication within MASs. 

Rasal presents a novel multi-agent communication framework named LLM Harmony~\cite{rasal2024llm}, which is inspired by the CAMEL model~\cite{li2023camel}, to enhance the problem-solving capabilities of Large Language Models (LLMs). The framework employs multiple LLM agents, each with a distinct role, engaged in role-playing communication. For their experiments, paper~\cite{rasal2024llm} focused on two metrics: arithmetic reasoning where math problems would be solved and common-sense reasoning. Each agent was given a distinct role to perform and their collaborative performance was measured. Their experiments validate the MAS framework's superior performance and adaptability, showcasing the potential of multi-agent systems in overcoming the limitations of individual LLMs.

AgentCoord~\cite{pan2024agentcoord} is a novel framework that has been designed to aid in the creation of coordination strategies for LLM-MASs. Liu et al.~\cite{pan2024agentcoord} proposed a three-stage generation method to transform user-provided goals into executable strategies. AgentCoord facilitates user comprehension by visually organizing the generated strategy and providing interactive features to explore alternative strategies. In a similar vein, AgentLite~\cite{liu2024agentlite} is another research based task-oriented planning framework that allows for the creation of LLM-MASs for researchers to study reasoning strategies, prototype novel LLM-MASs and reproduce agent benchmarks. The framework is validated through experiments conducted on complex reasoning tasks such as Retrieval-Augmented Q/A across a large volume of corpus and e-commerce tasks.

\subsubsection{Complex Workflows}
MASs have opened a pathway towards tackling more sophisticated scenarios in technical environments. Autogen, as explained in Section~\ref{sec:autogen} is one example of an LLM orchestration framework that leverages the power of LLMs to create autonomous agents~\cite{wu2023autogen} that can work together to solve complex scenarios and automate work-flows. Due to its capabilities, it can work as a research assistant, a programmer and code executor, a critic, reviewer and much more. Autogen is a representative of the current state-of-the-art capabilities of MAS frameworks that utilize the power of LLMs and are able to tackle work-flows in various domains. Besides auto-gen, there are a host of other MASs frameworks that enable complex workflows and event processing.

MetaGPT~\cite{hong2023metagpt} presents an innovative approach for leveraging large language models (LLMs) in multi-agent systems. This framework emphasizes the integration of human-like Standard Operating Procedures (SOPs) and a unique executable feedback mechanism to enhance the problem-solving capabilities and code generation quality of autonomous agents. It represents a specialized LLM application that focuses on automated software development. MetaGPT allows for the abstraction of software development roles for each agent (manager, developer, tester etc...) and automate the coding process in a collaborative manner. The experimental results highlighted MetaGPT's efficiency and robustness in real-world programming scenarios, confirming its task completion rate and reduced human revision costs. The primary limitations of the MetaGPT framework include its dependence on predefined SOPs, which might not be adaptable to all possible real-world scenarios. Unlike Autogen, MetaGPT is a specialized MAS LLM application and so is highly domain specific.

Similarly to MetaGPT, there are other MASs focused on the domain of software development. ChatDev~\cite{qian2024chatdev} is one such novel project that utilizes multi-agent communication to contribute to the design, coding, debugging and testing phases in a software development life cycle. Similarly, Feldt et al.~\cite{testingfeldt2023} presents a taxonomy of LLM-based testing agents that leverage conversational frameworks to automate software testing. This framework, named~\textit{SOCRATEST}, aims to provide varying degrees of autonomy to testing agents, enabling them to assist developers in the software testing process. The proposed system allows developers to specify high-level testing requirements, while the LLM-driven agents manage the detailed execution of tests, thus bridging the gap between testing expertise and domain knowledge. Extensive experiments were conducted to validate the effectiveness of the framework with various software testing tasks, such as unit testing. The experiments demonstrated that LLM-driven agents could generate detailed and actionable testing checklists, suggest relevant testing methodologies, and produce executable test code.

Other software related MASs have also attributed to ongoing research in this field. One common limitation among LLM-MASs was their inability to execute external tools (API and function calls) which limited complex use-cases from being automated via the use of these systems. Autogen~\cite{wu2023autogen}, Smurfs~\cite{chen2024smurfs} and ToolLLM \cite{qin2023toolllm} are examples of LLM-MASs that provide a framework that allows for the integration of external tool calls. Smurf~\cite{chen2024smurfs}, like Autogen, is a complex LLM-MAS framework that utilizes agents with specialized roles to break down complex prompts from the end-user, create an action plan, execute the created plan by utilizing external tools and finally verify and answer the prompt based on the preceding agents outputs. Unlike Smurf and auto-gen, ToolLLM~\cite{qin2023toolllm} is a framework designed to specifically enhance tool-use capabilities of LLM-MASs by enabling them to interact with a wide range of of APIs. The framework utilizes a fine-tuned version of the LLaMA model~\cite{Touvron2023LLaMAOA} to achieve its functionality. ToolLLM addresses this gap by be being trained on a comprehensive dataset called ToolBench~\cite{qin2023toolllm}, which includes 16,464 real-world APIs across 49 categories. The framework involves a three-stage construction process: API collection, instruction generation, and solution path annotation. The LLM-MAS is paired with a neural API retriever that recommends API calls based on the context provided. The framework was able to demonstrate remarkable performance in handling complex instructions and generalizing unseen APIs and exhibited performances comparable to state-of-the-art models at the time of writing.

Combining the principles and capabilities introduced by the aforementioned LLM-MASs frameworks that allow for code generation, execution, external tool calls combined with LLM multi-modal capabilities, exciting new pathways have been researched in the field of robotics. Papers~\cite{wang2024large, wang2024llmbased, kannan2024smartllm} highlight the usage of LLM-MASs and how the field of robotics can harness the advanced reasoning and language comprehension capabilities of state-of-the-art LLMs to create precise and functional action plans based on natural language instructions and external observations. Papers~\cite{wang2024large, wang2024llmbased} explore and demonstrate the capabilities of LLM-MASs for precise task planning and embodied tasks by utilizing vision models and combating LLM hallucinations. Similarly, Kannan et al.~\cite{kannan2024smartllm} proposes a framework titled \textit{Smart-LLM} that aims to convert high-level task inputs into a multi-robot task plan using a process that includes task decomposition, coalition formation, and task allocation. The framework is validated through extensive experiments in both simulation and real-world settings, demonstrating its effectiveness in generating coherent and logical multi-robot task plans. 

MASs have also been utilized in the field of cyber-security and networks. To demonstrate the capabilities of an MAS in this domain, Fasha et al.~\cite{owaspllm2024} proposes a MAS framework to tackle security risks defined in the Open Web Application Security Project (OWASP) Top 10. They focused on mitigating the OWASP Top 10 security risks for LLM applications using intelligent agents. Their proposed model leverages the AutoGen framework and Retrieval Augmented Generation (RAG) technologies to enhance the security of LLM deployments. The model employs autonomous agents to enforce security policies, validate inputs and outputs, and manage configurations. By integrating multiple intelligent agents working in a collaborative manner, the model can be seen to provide real-time security assessments and proactive countermeasures, enhancing data integrity, confidentiality, and service availability in LLM applications. Dandoush et al.~\cite{dandoush2024large} presents a comprehensive study on integrating Large Language Models (LLMs) with multi-agent systems (MAS) to enhance network slicing management and orchestration. The proposed framework leverages LLMs to translate user intent into precise technical requirements, map network functions to the infrastructure, and manage the entire slice life-cycle. The proposed framework facilitates collaboration across different administrative domains, overcoming challenges such as interoperability, security, resource coordination, and dynamic service orchestration.

The multi-modal capabilities of LLM have also proven to be a disruptive technology in the world of art and sciences. Models like GPT-4~\cite{openaigpt42024} and Google Gemini, have the ability to generate images based on textual input. Hence, LLM MASs have also been utilized in the art domain to enhance the designing process. Ding et al.~\cite{design2023} introduced DesignGPT, a multi-agent collaboration framework designed to enhance the product design process by integrating Large Language Models (LLMs) and generative AI tools. The system aims to simulate the roles within a design company, allowing human designers to collaborate with AI agents through natural language interactions. The system leverages the capabilities of GPT-4 for text generation and Stable Diffusion for image generation, providing design inspiration and aiding in the conceptual stage of product design. By integrating these tools, DesignGPT aims to bridge the gap between design thinking and machine thinking, enabling more effective and innovative design solutions 

\subsection{LLM-based Autonomous Agent: Future Trends and Gap Analysis}
Future research in LLM-based autonomous agents is likely to focus on improving the robustness and adaptability of both single and multi-agent systems. Enhancing the ability of agents to adapt to new tasks without extensive modification, developing more sophisticated communication protocols among agents and further integrating multi-modal LLMs to expand the capabilities and applications of these autonomous agents. 

\begin{figure}[t]
  \centering
   \includegraphics[width=\linewidth]{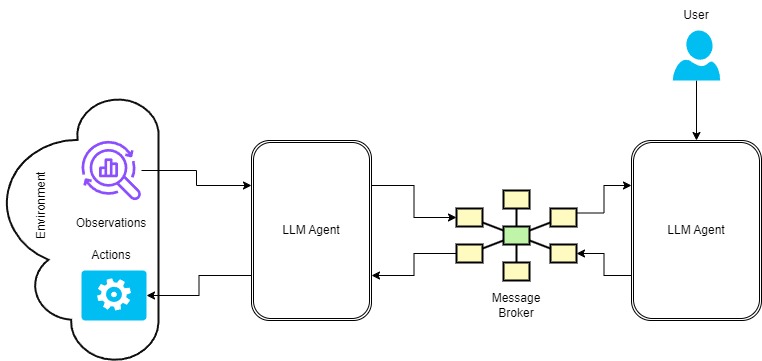}
   \caption[]{Proposed GenAI Broker Architecture - Adapted from~\cite{saleh2023pubsub} with Explicit Permission}
   \label{fig:message-broker}
\end{figure}

One such direction that is already being actively researched is related to augmenting the role of message brokers with LLM agents. Saleh et al.~\cite{saleh2023pubsub} introduces a novel framework for message brokers that are specially designed to cater to the needs of Generative AI (GenAI) applications by utilizing LLM agents. This approach is distinguished by its focus on scalability, efficiency, and the capacity for real-time data management, directly addressing the unique challenges posed by GenAI. The paper proposes utilizing an LLM agents and their reasoning ability to dynamically adjust to varying data loads and structures, ensuring that data can be processed and transferred without significant delays or bottlenecks. This adaptability is crucial for maintaining the performance and efficiency of GenAI applications, especially in scenarios where real-time processing and decision-making are essential. Figure~\ref{fig:message-broker} showcases a high level overview of a possible architecture as presented in the original paper~\cite{saleh2023pubsub}.

Another possible future direction that can utilize LLM agents is the proposal of a new neural pub/sub paradigm in paper~\cite{lovén2023ai}. The Neural Publish/Subscribe Paradigm aims integrates artificial intelligence directly into the communication layer, enabling a novel approach to managing AI workflows in distributed systems. It aims to leverage neural networks to facilitate efficient, many-to-many distribution of information, supporting dynamic learning and inference across the network. This can be supported by utilizing the power of LLM agents and presents an opportunity to introduce a more reactive and event-driven process, allowing various stages of the ML pipeline to act as subscribers.

\begin{figure}[t]
  \centering
   \includegraphics[width=0.8\linewidth]{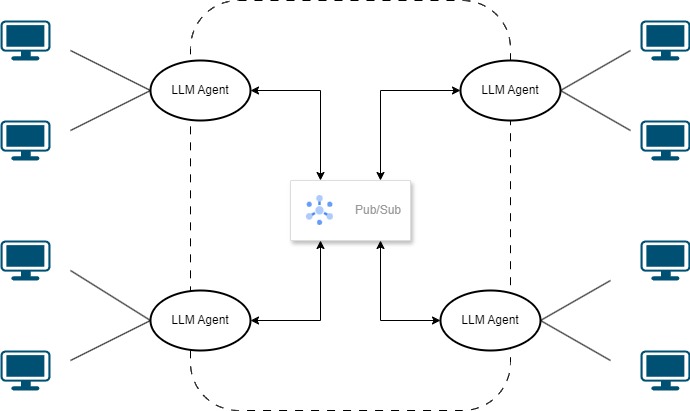}
   \caption[]{One possible architecture of a system adhering to the neural pub/sub paradigm}
   \label{fig:neural-pub-sub}
\end{figure}

Another possible future direction that can utilize LLM agents is the proposal of a new neural pub/sub paradigm (shown in Figure \ref{fig:neural-pub-sub}) in paper~\cite{lovén2023ai}. The Neural Publish/Subscribe Paradigm aims integrates artificial intelligence directly into the communication layer, enabling a novel approach to managing AI workflows in distributed systems. It aims to leverage neural networks to facilitate efficient, many-to-many distribution of information, supporting dynamic learning and inference across the network. This can be supported by utilizing the power of LLM agents and presents an opportunity to introduce a more reactive and event-driven process, allowing various stages of the ML pipeline to act as subscribers.

\section{System Design and Implementation}\label{sec:system-design}

LLM-empowered infrastructure promises to not only support dynamic analytics and decision-making but also enhance the overall efficiency and effectiveness of operations. The ability to provide seamless data flow and accessibility can empower users to respond swiftly to dynamic conditions. Considering such an objective, this section presents technical requirements, design principles, and the system design of the automated CEP pipeline employing LLM which can be utilized in modern infrastructures.

\subsection{Technical Requirements}\label{sec:requirements}
The primary objective of this system is to create a responsive and agile infrastructure that supports near real-time analytics and decision-making. This goal encompasses several key aspects:
\begin{itemize}
    \item \textbf{Timely Data Availability:} Ensuring that data is accessible when needed is crucial for real-time analytics and decision-making processes. The system must be capable of delivering data with minimal latency to facilitate prompt responses to emerging situations.
    
    \item \textbf{Data Accessibility:} The system must ensure that data is available at the right place, meaning it must support distributed environments where data needs to be accessed from various locations. This is particularly important for organizations with multiple branches or operations spread across different geographical regions.
    

    \item \textbf{Optimized Data Flow:} The system aims to maintain or optimize the flow of data across various stages of the data pipeline. This includes efficient data ingestion, processing, and retrieval processes that minimize bottlenecks and ensure smooth data transitions.

    \item \textbf{Low Latency:} Edge computing resources provide localized data processing, significantly reducing latency and enhancing the system's capability for real-time analytics and decision-making. Low latency is critical for applications that require immediate feedback, such as autonomous driving and real-time monitoring.

    \item \textbf{Scalability and Flexibility:} The data infrastructure must be scalable to handle increasing volumes of data and flexible enough to adapt to changing requirements. This includes the ability to integrate new data sources and support evolving analytical needs without significant reconfiguration.

    \item \textbf{Data Quality and Integrity:} Maintaining high data quality and integrity is essential. The system should include mechanisms for data validation, error detection, and correction to ensure that the data used for analytics and decision-making is accurate and reliable.
\end{itemize}

By achieving these objectives, the infrastructure can not only support dynamic analytics and decision-making but also enhance the overall efficiency and effectiveness of operations. The ability to provide seamless data flow and accessibility can empower users to respond swiftly to dynamic conditions.

\subsection{Design Principles}
\label{sec:design_principles}
The proof-of-concept presented in this thesis follows key principles aimed at ensuring the creation of a available and reliable system. These principles are as follows:

\begin{itemize}

    \item \textbf{Seamless Integration:} The solution is designed to integrate seamlessly with existing systems without causing breaking changes or compatibility issues. This principle ensures that the new system can be adopted smoothly, maintaining backward compatibility with existing software stacks and protocols. Such seamless integration minimizes disruption and facilitates smoother transitions to the new system.

    \item \textbf{Utilizing Existing Tools, Libraries, and Software:} The solution strategically utilizes existing open-source tools, libraries, and software's. By leveraging established resources, the design not only reduces development time and costs but also benefits from the robustness and reliability of well-maintained and widely-used components. This approach ensures that the system is built on a solid foundation of proven technologies.
\end{itemize}

\subsection{Overview of the System Architecture}\label{sec:overview_sys_arch}

\begin{figure}
    \centering
    \includegraphics[width=\linewidth]{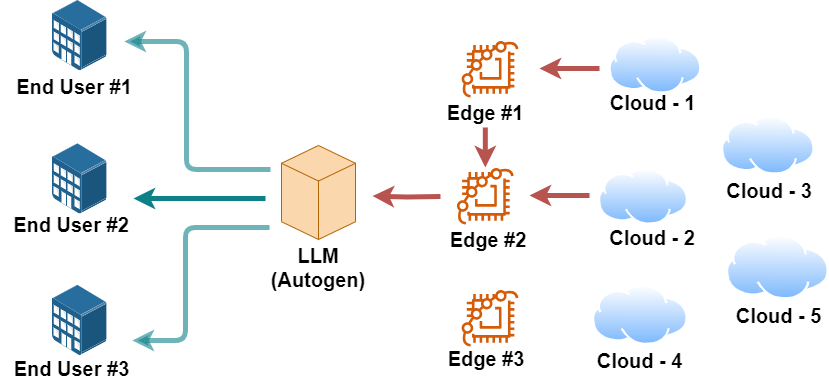}
    \caption{A High-Level System Overview}
    \label{fig:sys_overview}
\end{figure}

Figure~\ref{fig:sys_overview} illustrates the IoT data fabric architecture designed to create an autonomous complex event processing (CEP) pipeline. Central to this architecture is the Autogen Large Language Model (LLM) framework, which acts as the orchestration agent, coordinating data flows and processing tasks across various components. The system ensures seamless data flow, and accessibility, enabling efficient and intelligent decision-making. The diagram consists of three main components; the Central Coordination Unit, The End Users and Cloud and Edge Instances.

\textbf{Central Coordination Unit:}
At the core of the system is the Autogen LLM framework. This central coordination unit orchestrates the interactions between various components, managing agent routing, tool call allocation, and system integrity. Users interact with the system by querying the Autogen LLM agent, which autonomously orchestrates the necessary processes to fulfil the query. The framework leverages advanced GPT capabilities to understand user queries, determine the optimal data sources, and coordinate the processing tasks required to generate the desired outputs.
\begin{figure}
    \centering
    \includegraphics[width=0.8\linewidth]{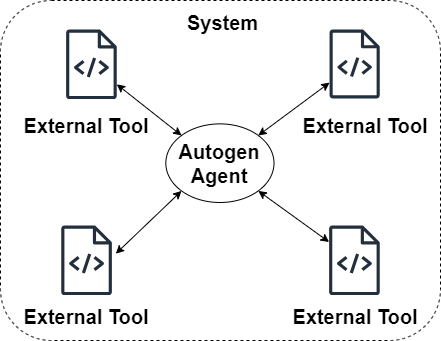}
    \caption{Bi-directional communication between an AutoGen agent and external tools}
    \label{fig:tools_autogen}
\end{figure}

A key feature of the AutoGen LLM framework is its capacity to integrate and leverage external tools and functions. AutoGen agents possess the capability to utilize external functions and tools to execute predefined workflows. This functionality is crucial for managing complex workflows that often require interaction with existing files, code bases, and functions rather than developing new code from scratch. Generating dynamic code can introduce breaking changes, increase latency, and pose additional security risks. By providing external tools to the AutoGen agent and customizing the agents to operate strictly within the boundaries of the provided code files, these drawbacks can be effectively circumvented.

The process of integrating external tools involves creating a bi-directional communication channel between the AutoGen agents and the external tools and/or code files. This interaction allows the agents to execute tasks using pre-existing, verified code, thereby enhancing reliability and security while reducing the risk of introducing errors or vulnerabilities. Figure~\ref{fig:tools_autogen} illustrates the communication flow between an AutoGen agent and external tools within the system.

\textbf{End Users:}
End users represent different domains that require data access and processing capabilities. They could be individual users, enterprises, or other entities that interact with the Autogen LLM framework to access the data needed for their specific domains. This interaction is facilitated through dedicated communication channels that ensure reliable and timely data delivery tailored to the end users' requirements. The users query the Autogen agents, which then orchestrate the data retrieval and processing operations transparently. One of the key design principles behind this POC is to minimize the complexity for end users in terms of system setup and understanding the internal mechanics of the AutoGen framework. 

The objective is to enable end users to seamlessly access relevant data based on their queries to the LLM agent without the need for extensive technical knowledge or configuration efforts. To achieve this, the POC is designed to abstract away the complexities involved in setting up connections and configuring the system. End users are not required to understand the intricate details of how the agents operate or how the tools and functions are integrated within the framework. This abstraction ensures that users can focus on their queries and tasks without being burdened by the underlying technicalities. The interaction with the system is streamlined such that the only requirement for the end users is to establish a communication channel with the LLM agent. Once this channel is established, users can issue queries and receive the necessary data through the LLM agent. The agent, in turn, handles all the back-end processes, including interaction with external tools and execution of predefined workflows. 

This user-centric approach not only enhances the usability of the AutoGen framework but also ensures that the system can be adopted by a broader audience, including those who may not have a deep technical background. By simplifying the user interaction and focusing on seamless data access, we demonstrates its potential as an effective tool for leveraging LLM capabilities in our use-case application (Video querying).

\textbf{Edge Devices and Cloud Instances:}
The architecture includes cloud instances that host computational power to process data (e.g: they host the LLM). A cloud instance can perform complex data processing, storage, and analytics tasks. The Autogen LLM framework can coordinates these instances via function calls due to tool-backed agents, ensuring efficient distribution and task execution. This distributed cloud architecture enhances the system's ability to handle large volumes of data and complex processing workloads. In this POC, the cloud instances represent two aspects of the system; the GPT-4o model where the videos are analyzed, and the cameras that are the source of the footage (camera-1, camera-2 etc...).

Edge resources are deployed near the data sources. These resources handle preliminary data processing tasks close to the data generation points, reducing latency and bandwidth usage. By performing initial filtering, aggregation, and analysis, edge devices enhance the system's responsiveness and offload the central and cloud processing units. Edge computing is particularly beneficial for applications where low latency is critical. In the context of this POC, the edge instances represent the message brokers that forward frames from the "cameras" to the Autogen agents. The frames are assigned to the relevant topics, and the autogen agents subscribe to the topic relayed via user prompts to consume the frames needed to be processed. The agent then forwards the frames to the GPT model hosted on the cloud for processing details based on the user prompt.

\subsection{Complete Data Flow}\label{sec:data-flow}

\begin{figure}
    \centering
    \includegraphics[width=\linewidth]{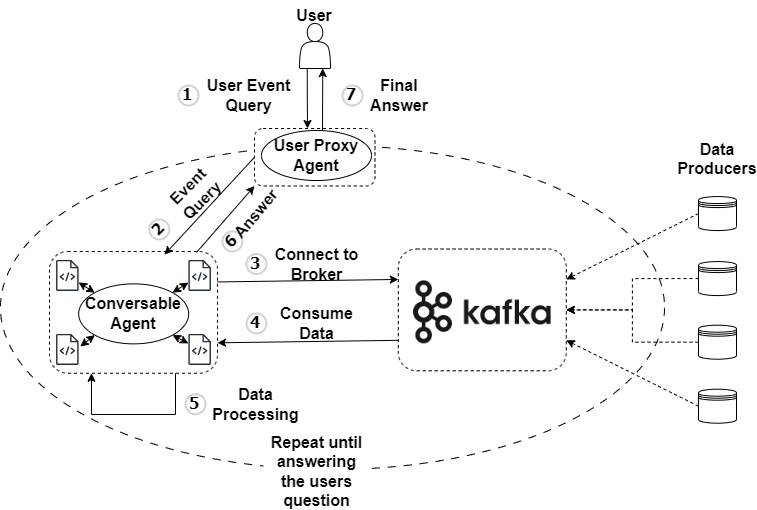}
    \caption{Complete Flow representing the sequence of steps taken by the system to resolve an event query issued by the user}
    \label{fig:comp-data-flow}
\end{figure}

Figure~\ref{fig:comp-data-flow} showcases the sequences of steps that take place when an end user interacts with the system. The end user initiates the interaction by posing a query, such as “Are there any red vehicles visible in camera 1, camera 2, camera 3, camera 4, camera 5?”, marked by \textbf{Step 1} to the User Proxy Agent. The scenario depicts a two-agent system for brevity. An addition of another agent would function as a conversable agent that would coordinate with the tool-backed agent with dynamic prompts, adding more nuance to the workflow. In this two-agent system, The User Proxy Agent manages and coordinates with the Conversable tool-backed Agent. The agents are also able to capture and retain valuable context gained during the workflow regarding the user's queries and their corresponding responses. Autogen allows this gained knowledge to be shared across the system with other agents which empowers them to give more informed and relevant responses.

In this orchestrated process, the Conversable Agent will connect to the Kafka broker and consume relevant data streams to gather information from the camera specified in the prompt (camera 1-5) \textbf{(steps 2 and 3)}. For instance, during data retrieval, the Conversable Agent will subscribe to the topic "camera-1", consuming the latest frames from that topic \textbf{(Step 4)}. After retrieving the data, the Conversable Agent will coordinate with a tool-backed agent that will process it using the tools it has been provided and perform any additional calculations or transformations needed to answer the user's query \textbf{(step 5)}. Once the data is processed, the Conversable Agent will interpret the results and provide an answer \textbf{(steps 6 and 7)} for the user prompt. For example, the Conversable Agent might say, “There are red cars present in camera 1 and camera 3.” Bringing this intricate process full circle, the User Proxy Agent relays to the user the concluding answer.

If at any point there is an exception—either an error in data retrieval or processing, the User Proxy Agent redirects the issue back to the Conversable Agent for resolution. If the issue cannot be resolved, the user is informed that an unexpected error has taken place. The end user can also be prompted for additional queries for further clarification. Hence, steps 3 to 6 can be repeated multiple times, depending on the prompts of the user. The entire workflow of multi-agent interaction and conversation is managed via Autogen, with the user only responsible for providing Q\&A prompts.

This detailed workflow demonstrates the efficiency and capability of the Autogen LLM framework in orchestrating an autonomous complex event processing pipeline, ensuring accurate and timely responses to user queries. By leveraging the advanced orchestration capabilities of the Autogen LLM, the system is able to ensure efficient data flow, processing, and decision-making. This architecture not only supports the scalability and flexibility required for dynamic IoT environments due to its modularity but also provides the necessary redundancy and fault tolerance to maintain continuous and reliable operations. The integration of advanced AI capabilities within the Autogen framework further enhances the system's ability to adapt and respond to evolving data and user requirements. 

One important point to note is that both synchronous and asynchronous settings are supported in our PoC. Users (i.e. subscribers) may use our PoC to make event queries and receive answers both in the real-time. They may also subscribe to their queries, and matching events will be stored in buffer which users can receive asynchronously.

\section{Experiments}\label{sec:exp}
This section introduces robustness and performance of the proposed system. The experiments focus on measuring the system's efficiency with varying configurations and input complexities. The two key experiments are detailed: the first experiment assesses the system's performance as the number of Autogen agents increases, while the second experiment evaluates the impact of increasing input video complexities on the system's performance.

\subsection{Evaluation Criteria and Methodology}
\label{sec:evaluation-criteria}
In this section, we introduce the evaluation criteria used to assess the accuracy and quality of video descriptions generated by the system. Given that our architecture employs multiple agents, each with specialized roles, it is crucial to ensure that the system’s video analysis performance remains high and consistent, reflecting the individual contributions of each agent. The evaluation criteria utilizes the same methodology introduced in the Video-ChatGPT paper~\cite{maaz2023videochatgpt}. This criteria was chosen to provide a comprehensive evaluation of the descriptions, ensuring that they accurately represent the content of the videos and convey the necessary information effectively. Each criterion chosen by the Video-ChatGPT~\cite{maaz2023videochatgpt} team is detailed below, along with the methodology for scoring each video level.

\begin{itemize}
    \item \textbf{Correctness of Information:} 
This criterion measures the extent to which the information provided in the video description aligns with the actual content of the video. It evaluates whether the key details and events described are factually accurate and correctly portrayed. A high score in this category indicates that the description reliably represents the video's content without introducing inaccuracies. The score for this criterion is calculated by comparing the presence and accuracy of key details in both the human-generated and LLM-generated descriptions.

\item \textbf{Detailed Orientation:}
Detailed Orientation assesses the level of detail included in the description. This criterion looks at whether the description provides sufficient granularity, covering specific aspects such as objects, actions, and settings. A detailed-oriented description helps in painting a clear and vivid picture of the video content, facilitating better understanding and visualization. The score for Detailed Orientation is calculated by evaluating the comprehensiveness of the details provided in the LLM-generated description compared to the human-generated one.

\item \textbf{Contextual Understanding:}
Contextual Understanding evaluates the description’s ability to correctly interpret and convey the broader context of the video. This includes understanding the setting, the relationships between different elements within the video, and the overall scene. High scores in this category signify that the description not only mentions key elements but also places them within a coherent and accurate context. The score is determined by assessing how well the LLM-generated description captures the context compared to the human description.

\item \textbf{Temporal Understanding:}
Temporal Understanding focuses on the description's capability to accurately depict changes and sequences over time within the video. This criterion examines whether the description correctly narrates the progression of events, transitions between scenes, and any temporal shifts that occur. Effective temporal understanding ensures that the description reflects the dynamic nature of the video accurately. The score for Temporal Understanding is calculated by comparing how well both descriptions capture the progression and changes over time.

\item \textbf{Consistency:}
Consistency measures the uniformity and stability of the description throughout the entire video. A consistent description maintains a steady level of detail, accuracy, and contextual relevance across different parts of the video. This criterion ensures that the description does not fluctuate in quality or coherence, providing a reliable and continuous narrative from start to finish. The score is calculated by evaluating the consistency of the LLM-generated description in comparison to the human description.
 
\end{itemize}

\subsubsection{Scoring Methodology}
Each video level is evaluated against the five criteria listed above. For each criterion, a score ranging from 0 to 1 is assigned based on the description's performance:
\begin{itemize}
 \item  1.0 indicates that the description meets the criterion excellently, with no significant issues.
  \item  0.9 indicates that the description meets the criterion well but has minor issues.
 \item  0.8 indicates that the description meets the criterion satisfactorily but with some noticeable issues.
 \item  0.7 or lower indicates more significant issues in meeting the criterion.
\end{itemize}
This scoring methodology provides a quantitative measure of the description’s overall quality, highlighting both strengths and areas for improvement.

\subsection{Experiment 1: Measuring performance with increasing number of Autogen Agents}
\label{exp:Increase_agents}
The first experiment aimed to investigate the impact of the number of Autogen agents on the system's workflow execution. To maintain consistency, the same video and prompt were used across all trials to ensure that video and prompt complexity remained constant and did not influence system performance. For an in-depth analysis of how prompt and video complexity affect system performance, refer to Section \ref{exp:experiment2}. The performance was evaluated by incrementally increasing the number of agents and observing the system's ability to handle tasks.

\begin{itemize}
    \item \textbf{Minimum Number of Agents 2:}
    \newline
    The system requires a minimum of two agents to complete the workflow. The user proxy agent is responsible for acting on behalf of the user in the system, while the engineer agent is tasked with calling the appropriate tools to complete the workflow. Although the system could initiate the workflow, it was unable to continue beyond the first prompt, leading to termination. The Engineer agent can only execute function calls and return their results but cannot engage in further conversation beyond the initial single prompt. This indicates that two agents are insufficient for maintaining the workflow.

    \item \textbf{Increase Number of Agents to 3:} 
    \newline
    In this scenario, the user-proxy agent, responsible for acting on behalf of the user, and the Engineer agent, which decides which functions to call and has access to them, retained their previous roles. A new third agent, named the "Reflection-agent," was introduced to converse with the user (via the user-proxy agent) and to add question and answer (Q/A) functionality to the system by utilizing the results provided by the Engineer agent's function calls.
    
    With the introduction of three agents, the system enabled additional question and answer (Q/A) interactions concerning the analyzed video. However, the context was limited to the initial video analysis, requiring the Reflection-agent to instruct the Engineer agent to reanalyze the video for further processing. This configuration improved continuity but still exhibited limitations in maintaining comprehensive context.

    \item \textbf{Increase Number of Agents to 4:} 
    \newline
    A fourth agent, named "Engineer-2," was introduced to the system. This new agent was tasked with executing a subset of functions during the workflow, while the original "Engineer" agent was updated to call the remaining tools. The function calls were thus divided between the two agents, with the "Reflection" agent retaining its role from the previous setup.
    
    The addition of the fourth agent allowed for more granular control over function calls by assigning specific roles to each agent. However, this configuration increased the overall system latency due to the more frequent communications among the agents. Despite the increased latency, this setup offered a more optimized workflow and improved function execution management.
    
\end{itemize}

\captionsetup[table]{justification=raggedright,singlelinecheck=off}
\begin{table*}[h!]
\centering
\begin{tabular}{|c|c|c|}
\hline
\textbf{Number of Agents} & \textbf{Average Latency (seconds)} & \textbf{Average Agent Overhead (seconds)} \\
\hline
2 & 5 - 8 & 1 - 2 \\
\hline
3 & 12 - 16 & 8 - 10 \\
\hline
4 & 22 - 25 & 12 - 16 \\
\hline
\end{tabular}
\caption{Average Latency and Agent Overhead for Different Numbers of Agents}
\label{tab:number_of_agents}
\end{table*}

Table \ref{tab:number_of_agents} presents a summary of the system performance as the number of agents was increased from 2 to 4, including the average latency range measured for each configuration measured over repeated experimentation. For the purposes of consistency, the video and prompt used to evaluate the performance as the number of agents increased were kept consistent. With only two agents, the system exhibited minimal latency between 5-8 seconds. Introducing a third agent increased the average latency to a range between 12-16 seconds. The inclusion of a fourth agent further extended the average latency to a range of 22-25 seconds. 

\begin{figure}
    \centering
    \includegraphics[width=0.85\linewidth]{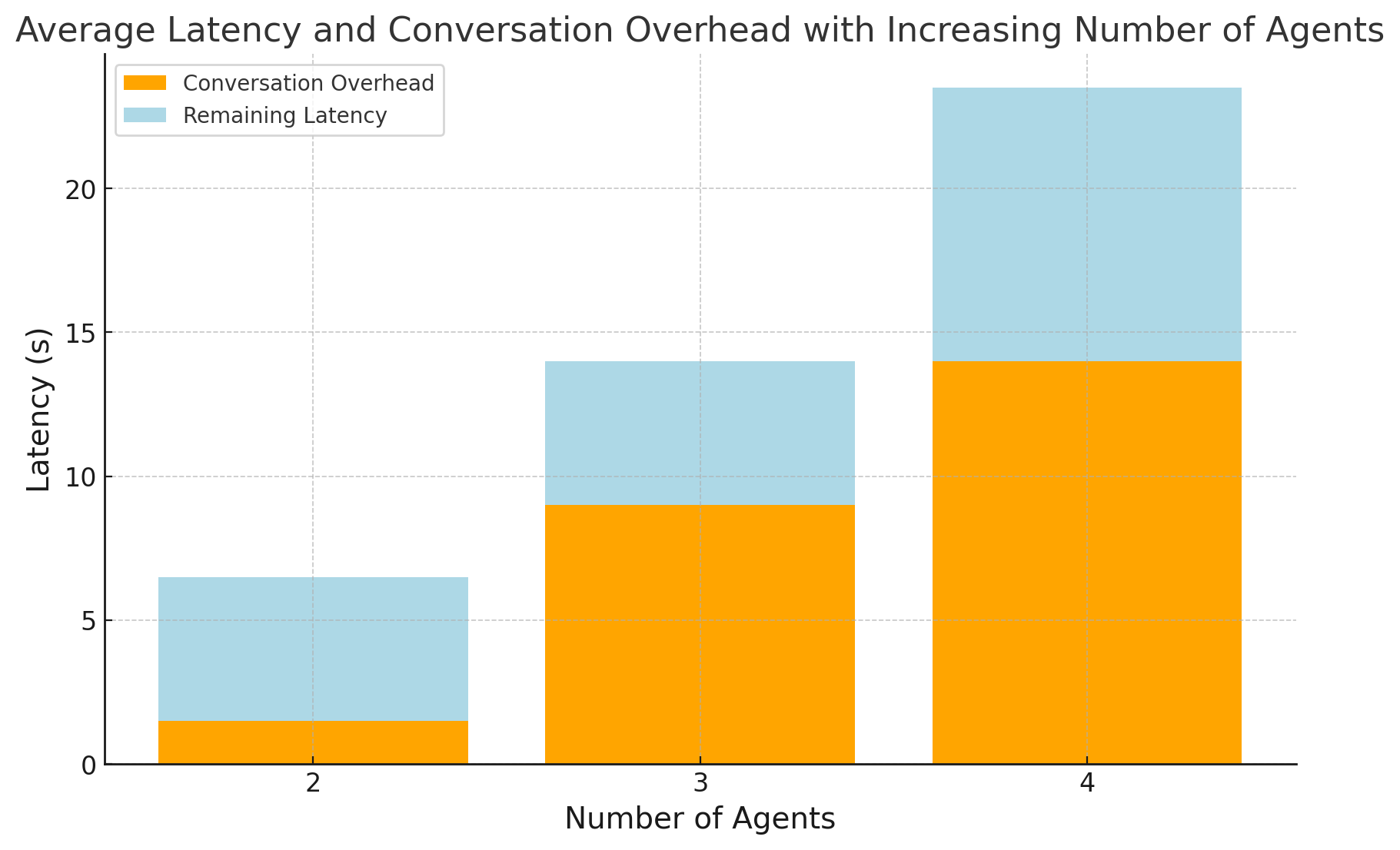}
    \caption{The effect on system latency with increasing number of agents}
    \label{fig:increasing_agents_latency}
\end{figure}

As shown, the average latency increases with the addition of more agents. Figure \ref{fig:increasing_agents_latency} showcases the results in a graphical format. We can observe that as the number of agents in the system increase, they contribute significantly more to the overall increase in the system latency due to the increasing overhead in conversations taking place among the agents themselves. 

It was observed that adding more than four agents led to excessive and redundant communications, further exacerbating latency without significant improvements in functionality. Thus, the optimal configuration appears to be four agents, balancing enhanced capabilities with manageable latency for this specific use-case. More complex use cases that utilize multiple tools and function calls would benefit from increasing the number of agents, with each agent managing a subset of tools and functions to call. However the system latency will need to be compromised in such a scenario.

\subsubsection{Speaker Selection Optimization}
\label{subsec:speaker-select}

\begin{figure}
    \centering
    \includegraphics[width=0.8\linewidth]{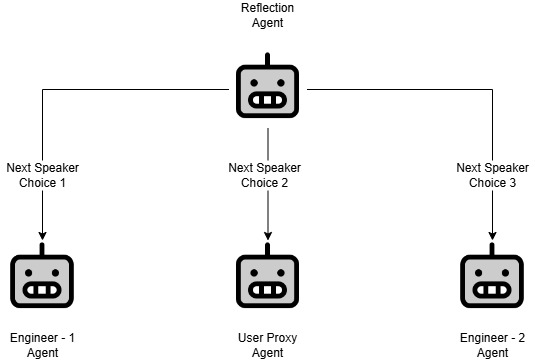}
    \caption{Next possible speaker choice after the reflection agent in a four-agent system}
    \label{fig:speaker-selection-graph}
\end{figure}

As the number of agents continues to increase, determining which agent is in charge at any given point in the workflow becomes a challenge. Autogen \cite{wu2023autogen} provides four configurations to determine the "speaker" agent: Auto, Manual, Round-Robin, and Random. Figure \ref{fig:speaker-selection-graph} showcases a decision-making scenario in a four-agent conversation, where the next speaker must be determined after the reflection agent.

The "Manual" configuration would make the system non-autonomous, requiring the user to understand the inner complexities and workings of the tools in the system and to manually guide the workflow. The "Random" configuration would result in random agents taking control, potentially sabotaging the workflow and rendering the system unusable.

Increasing the number of agents, while utilizing "Auto" or "Round-Robin" speaker selection results in a more complex scenario wherein the work-flow becomes difficult to manage or control. With two agents, both selection methods result in a normal two-agent conversation, indicating similar effectiveness. With three agents, both methods still maintain relatively accurate speaker selection. However, as the number of agents increases to four, the accuracy of speaker selection diverges. The Auto Selection method, which is context-aware, becomes less accurate, while the Round-Robin Selection method shows a notable decrease in accuracy, reflecting its limitations in managing a larger number of agents. This comparison highlights the importance of selecting an appropriate speaker selection strategy based on the number of agents and the specific requirements of the task to maintain system performance and accuracy.

\begin{figure}
    \centering
    \includegraphics[width=\linewidth]{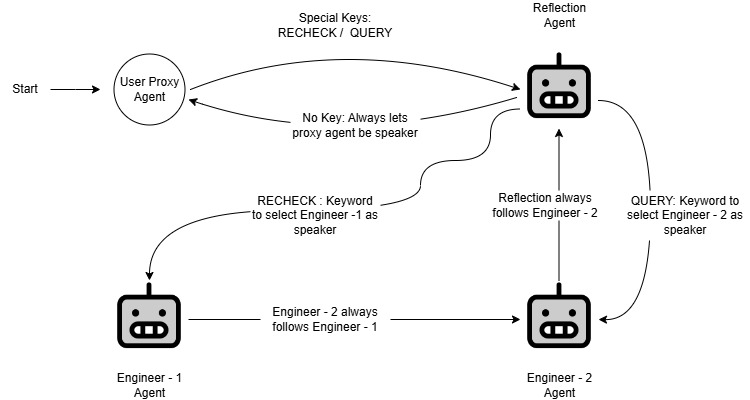}
    \caption{An optimized speaker selection method in a four-agent system}
    \label{fig:optimized_speaker}
\end{figure}

Figure \ref{fig:optimized_speaker} illustrates a customized, structured speaker selection process in a four agent system. The process begins with the User Proxy Agent, which initiates the task and continues conversing with the reflection agent unless a specific keyword is detected. The Reflection Agent evaluates the context and any special keywords, such as "RECHECK" or "QUERY," to determine the next speaker. If no keyword is detected, the User Proxy Agent and the Reflection agent continue with a normal conversation. If the "RECHECK" keyword is detected, the Reflection Agent selects Engineer-1 Agent as the next speaker, who then transitions the task to Engineer-2 Agent upon completion. If the "QUERY" keyword is detected, the Reflection Agent selects Engineer-2 Agent directly. After Engineer-2 Agent completes its task, it transitions back to the Reflection Agent. This decision-making process ensures that tasks are dynamically assigned to the most suitable agent, optimizing the system's performance and responsiveness. This structured approach allows for flexible and efficient task management within the system, ensuring seamless operation across various scenarios.

\subsection{Experiment 2: Evaluating performance with increasing video complexities}
\label{exp:experiment2}
The second experiment focused on evaluating the system's performance under varying levels of video input complexities, with a particular emphasis on the impact of video content complexity on system latency and accuracy. 

This experiment investigated the system's performance in analyzing videos with varying levels of content complexity. The aim was to assess how the intricacy of video data influences processing efficiency and the system's ability to provide accurate and comprehensive responses.

In our experiments, we analyzed the impact of video complexity on the performance of the system under various conditions. To systematically assess this, we categorized the videos into five distinct complexity levels. Each level is defined based on specific characteristics of the video content, such as the number of vehicles, their movement patterns, and the environment's complexity. The detailed descriptions for each video complexity level are provided in Table \ref{tab:vid_complexity} below. This classification allowed us to evaluate the system's performance comprehensively across a range of real-world scenarios. 

\begin{table}[h!]
\centering
\begin{tabular}{|l|p{4.6cm}|}
\hline
\textbf{Video Complexity Level} & \textbf{Description} \\
\hline
Level 1 \footnotemark[7]  & A video where only a single car is visible in an uncluttered environment. \\
\hline
Level 2 \footnotemark[8] & A video where a few cars pass through in an uncluttered environment. \\
\hline
Level 3 \footnotemark[9] & A video where multiple vehicle types are travelling in opposing directions at any given time. \\
\hline
Level 4 \footnotemark[10] & A video where vehicles are seen from an aerial view travelling in multiple directions in a cluttered environment. \\
\hline
Level 5 \footnotemark[11] & A sped-up video where multiple vehicles can be seen travelling in a cluttered environment with bright lights. \\
\hline
\end{tabular}
\caption{Descriptions for Different Video Complexity Levels}
\label{tab:vid_complexity}
\end{table}
\footnotetext[7]{https://www.youtube.com/watch?v=mX1Ju5Hvlaw}
\footnotetext[8]{https://www.youtube.com/watch?v=K6xsEng2PhU}
\footnotetext[9]{https://www.youtube.com/watch?v=zOq2XdwHGT0}
\footnotetext[10]{https://www.youtube.com/watch?v=DuUaCwM-Q10}
\footnotetext[11]{https://www.youtube.com/watch?v=XgqTf1f-5Hw}

\begin{figure}[ht]
    \centering
    \begin{minipage}{0.3\textwidth}
        \centering
        \includegraphics[width=\linewidth]{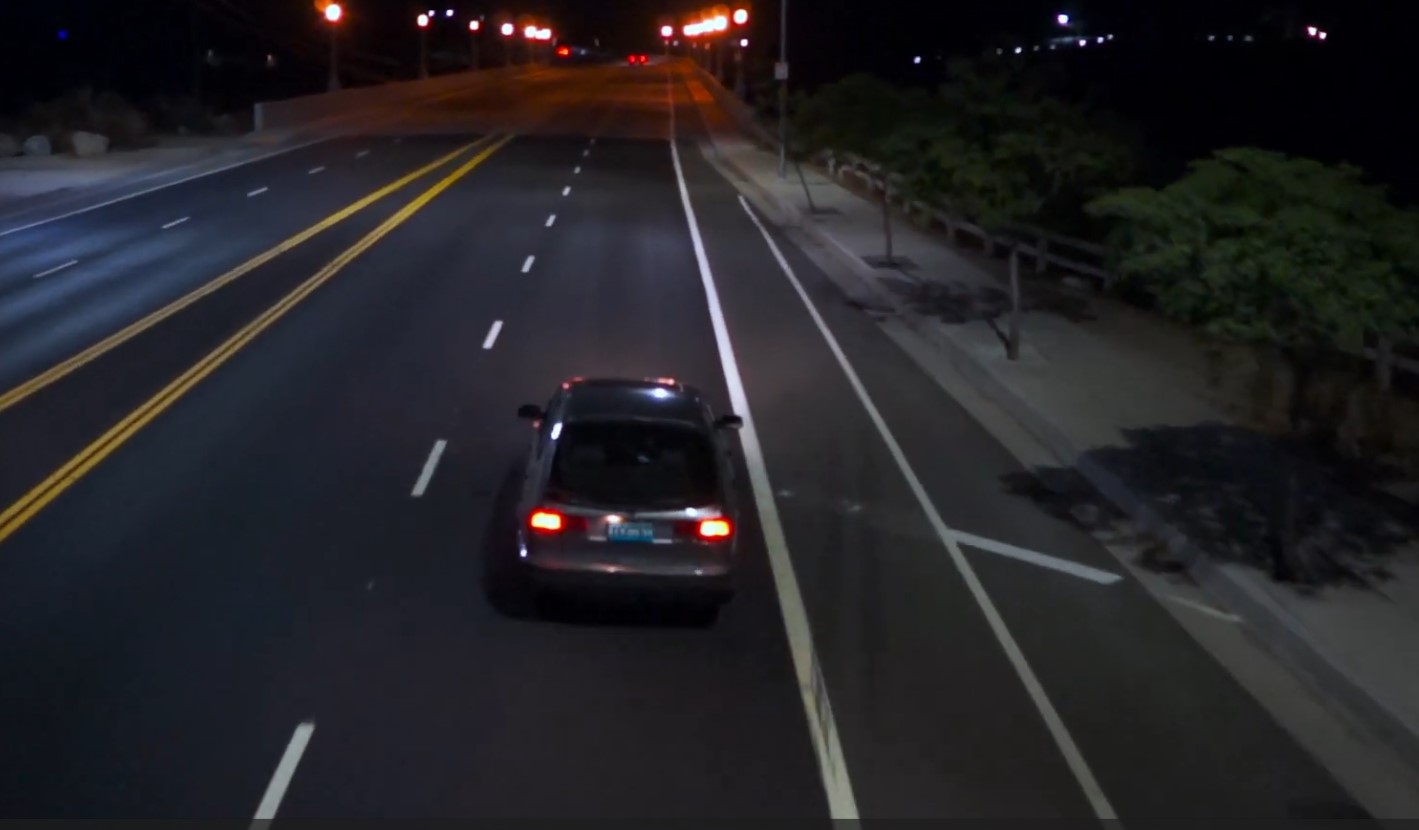}
        \caption{Level-1 Complexity}
        \label{Video-1}
    \end{minipage}\hfill
    \begin{minipage}{0.3\textwidth}
        \centering
        \includegraphics[width=\linewidth]{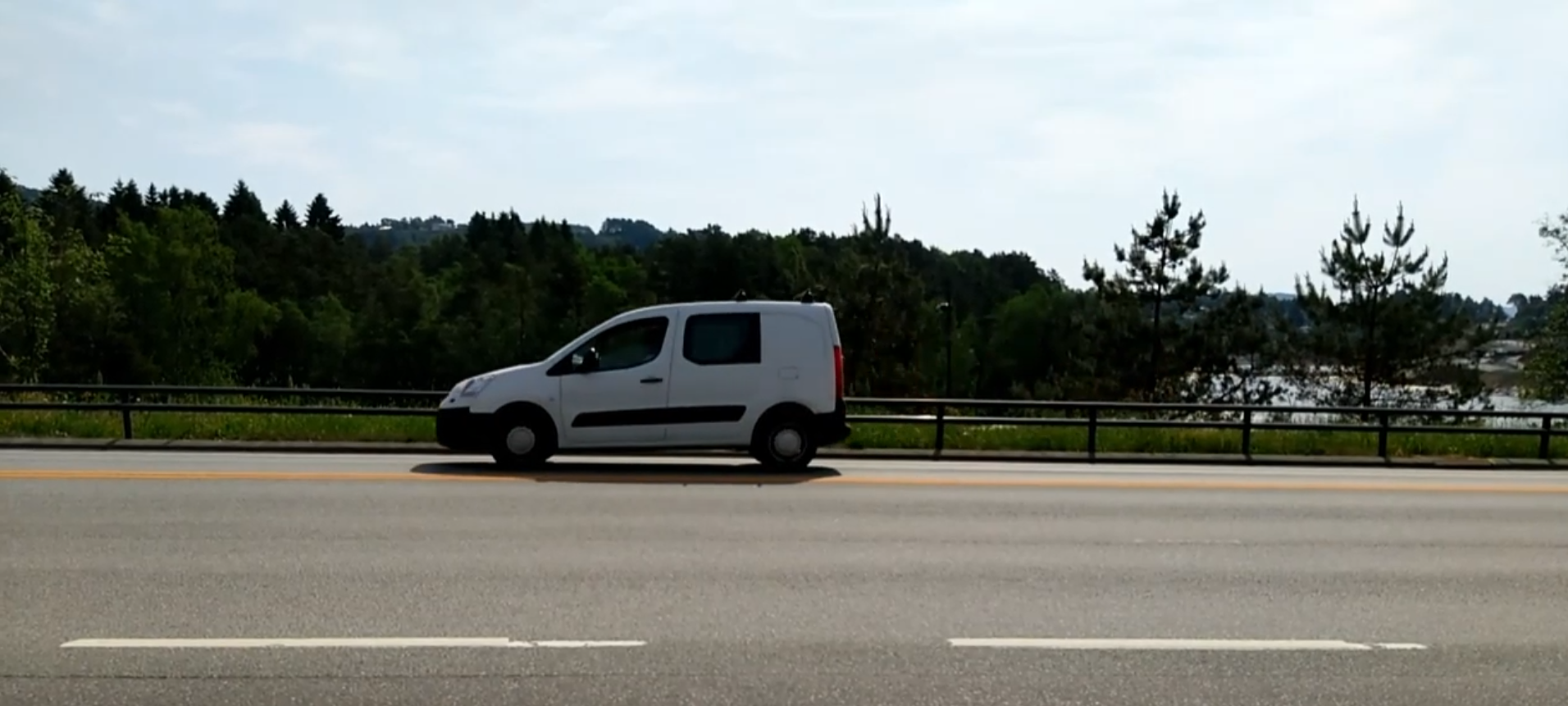}
        \caption{Level-2 Complexity}
        \label{Video-2}
    \end{minipage}\hfill
    \begin{minipage}{0.3\textwidth}
        \centering
        \includegraphics[width=\linewidth]{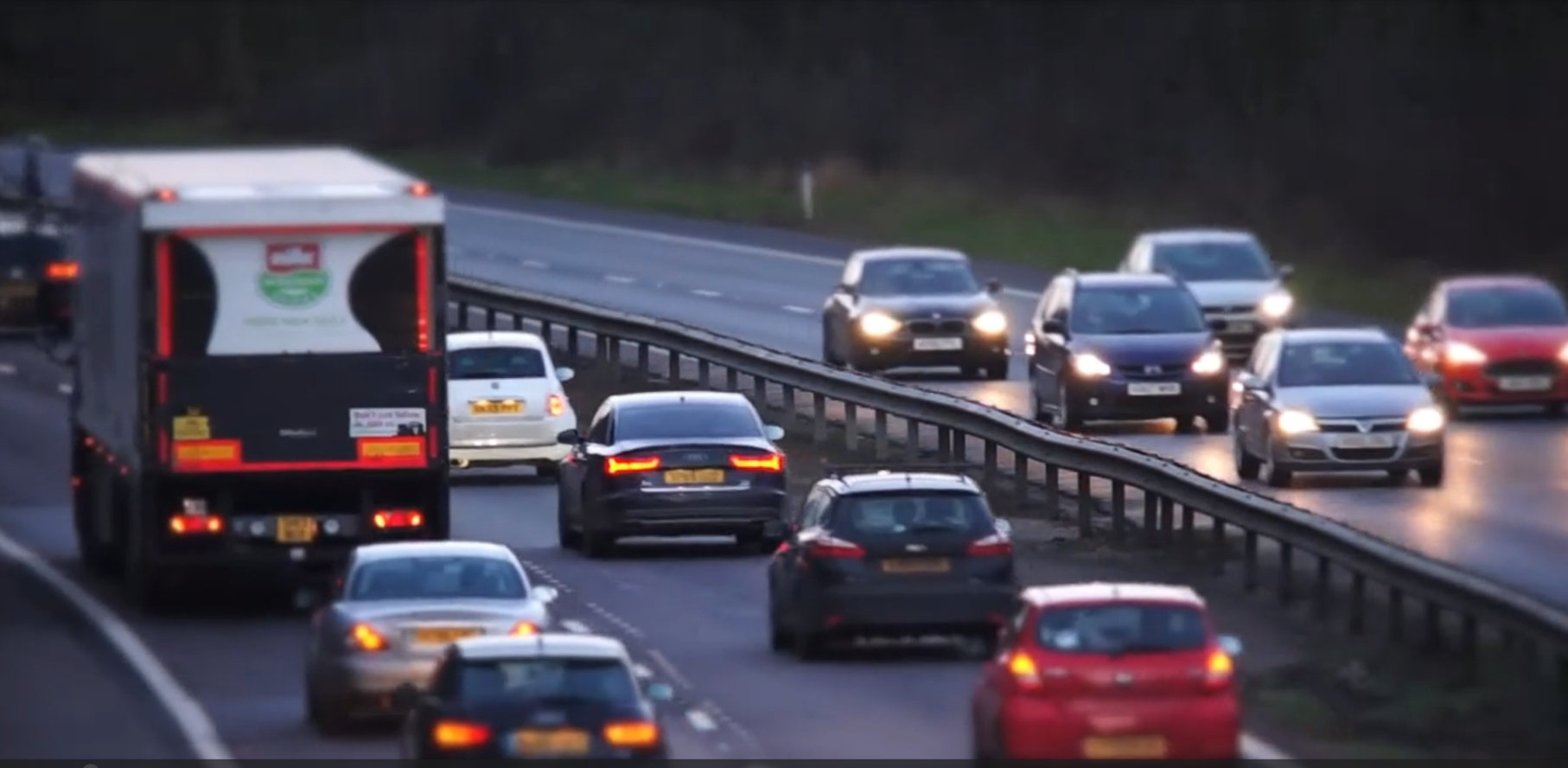}
        \caption{Level-3 Complexity}
        \label{Video-3}
    \end{minipage}
\end{figure}

\begin{figure}[ht]
    \centering
    \begin{minipage}{0.45\textwidth}
        \centering
        \includegraphics[width=\linewidth]{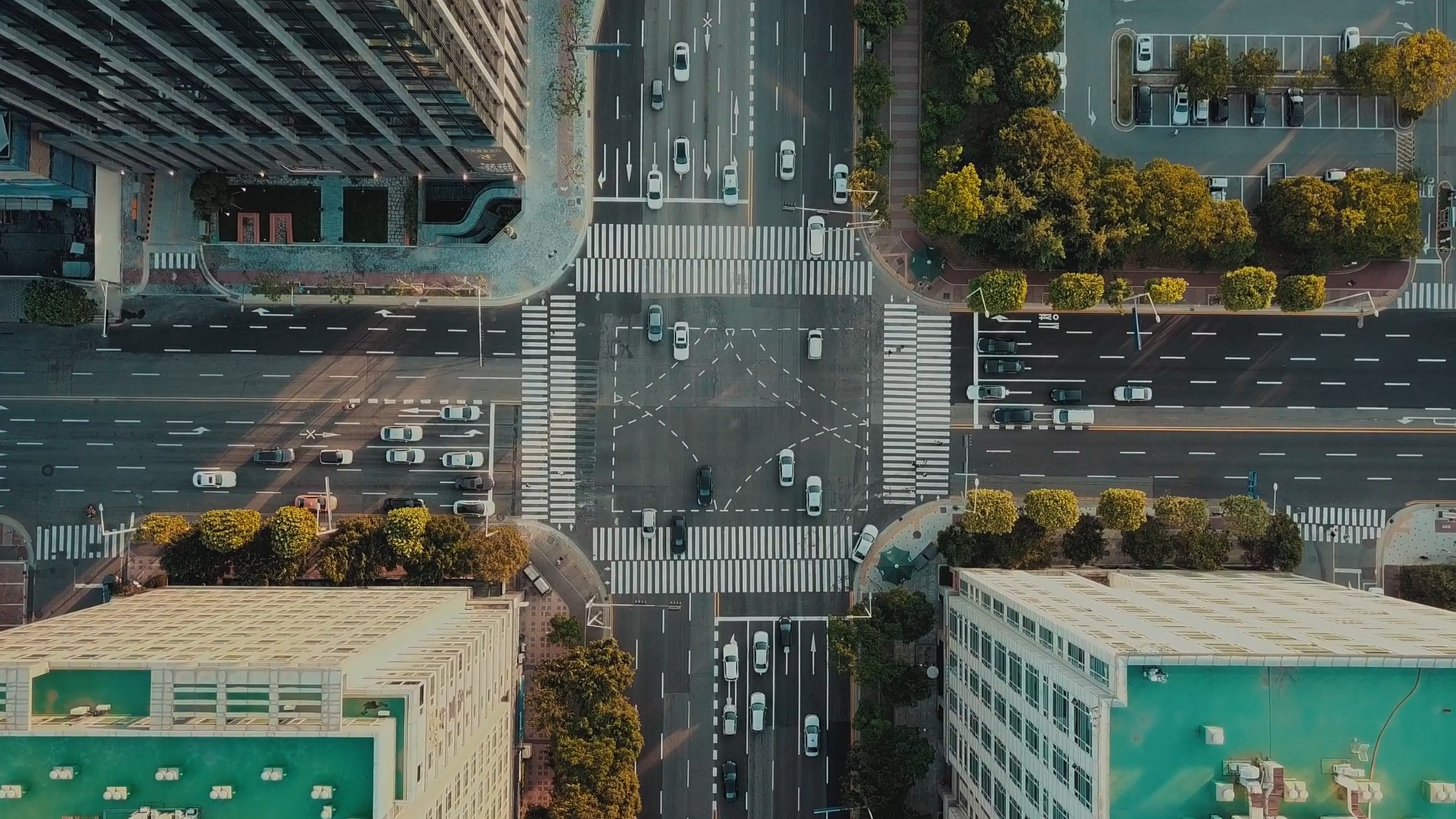}
        \caption{Level-4 Complexity}
        \label{Video-4}
    \end{minipage}\hfill
    \begin{minipage}{0.45\textwidth}
        \centering
        \includegraphics[width=\linewidth]{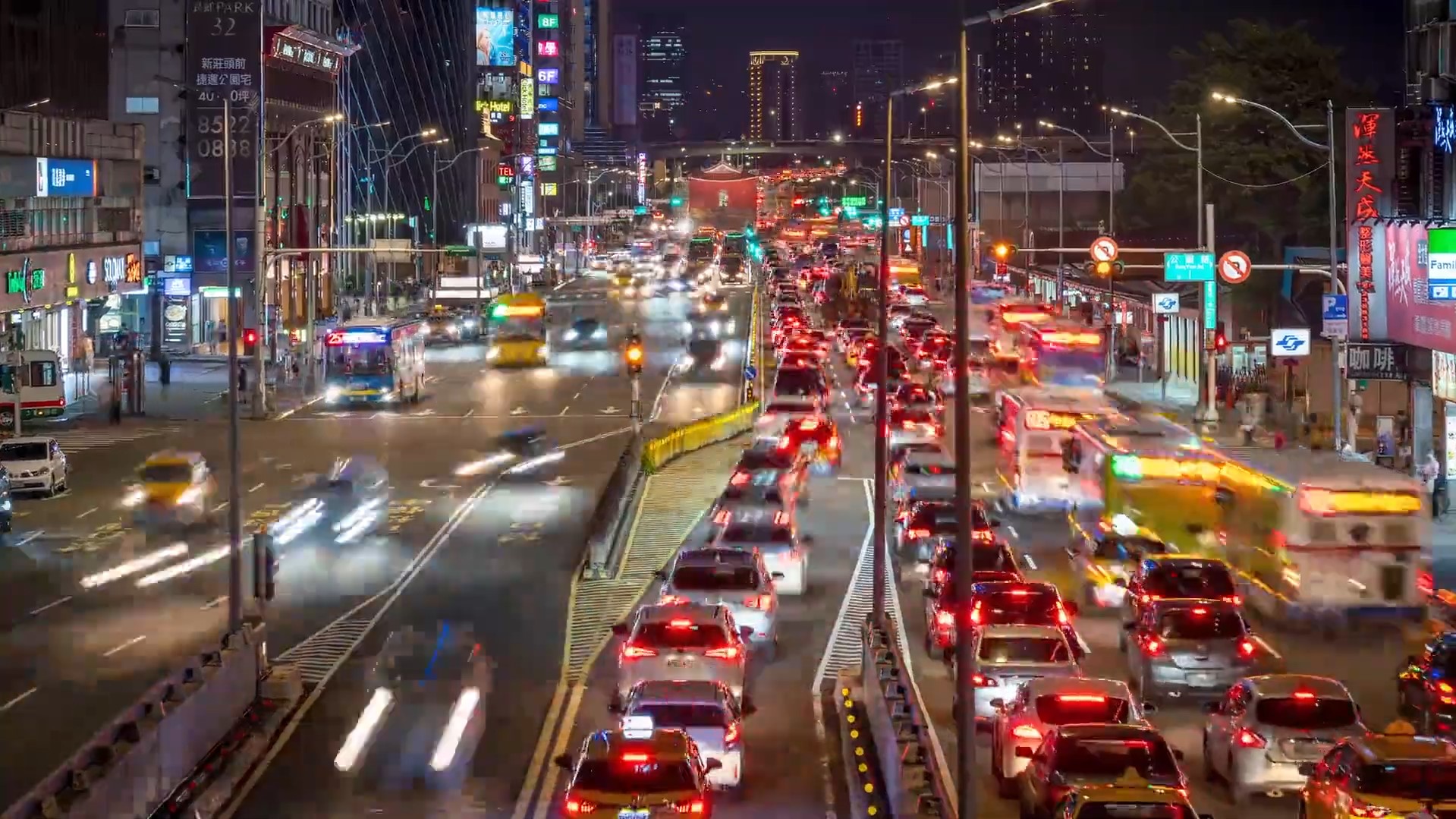}
        \caption{Level-5 Complexity}
        \label{Video-5}
    \end{minipage}
\end{figure}

Figures \ref{Video-1} to \ref{Video-5} illustrate these different levels of complexity, showcasing various scenarios from a single car in an uncluttered environment to multiple vehicles traveling in a cluttered setting with bright lights and different environmental factors.

\vspace{10pt}

To evaluate the performance of the system with increasing complexity levels as identified in Table \ref{tab:vid_complexity}, we utilized the methodology introduced in Section \ref{sec:evaluation-criteria}. Table \ref{all-eval-scores} summarizes the evaluation scores based on the introduced criteria. The evaluation is based on the five key criterion's: Correctness of Information, Detailed Orientation, Contextual Understanding, Temporal Understanding, and Consistency. Each criterion is scored on a scale from 0 to 1, where 1 indicates excellent performance and 0 indicates poor performance. To maintain consistency, we utilized a two-agent system so as to evaluate one complete work-flow cycle and to minimize internal agent conversation overhead that contribute significantly to the latency as described in Section \ref{exp:Increase_agents}.

\begin{table}[h!]
\centering
\begin{tabular}{>{\raggedright\arraybackslash}p{1.5cm}*{5}{c}}
\toprule
\textbf{Criteria} & \textbf{Level 1} & \textbf{Level 2} & \textbf{Level 3} & \textbf{Level 4} & \textbf{Level 5} \\
\midrule
Correctness of Information & 0.8 & 0.8 & 0.9 & 0.7 & 0.6 \\ \hline
Detailed Orientation & 0.7 & 0.7 & 0.8 & 0.6 & 0.5 \\\hline
Contextual Understanding & 0.9 & 0.9 & 0.9 & 0.8 & 0.7 \\\hline
Temporal Understanding & 0.8 & 0.8 & 1.0 & 0.7 & 0.6 \\\hline
Consistency & 1.0 & 1.0 & 1.0 & 1.0 & 0.9 \\
\bottomrule
\end{tabular}
\caption{Evaluation Scores for Each Video Level Complexity (@1080p resolution)}
\label{all-eval-scores}
\end{table}

Table \ref{all-eval-scores} consolidates the evaluation scores for each video level based on the evaluations and criteria, providing a clear and detailed assessment of the performance of the descriptions provided by the system in comparison to human-generated descriptions. 

We can observe from the evaluation scores that less complex videos featured straightforward and easily identifiable elements. For example, Level 1 and Level 2 videos, which depicted a single or handful of moving objects or simple scenes with minimal background activity, allowed the system to perform efficiently. The scores for these levels were generally high, with Level 1 and Level 2 each receiving scores of 0.8 for Correctness of Information and 0.7 for Detailed Orientation, indicating strong performance but with some room for improvement in detail and accuracy.

In contrast, more complex videos presented intricate scenarios with multiple moving objects, varied backgrounds, and dynamic interactions. For instance, the Level 3 video, featuring multiple types of vehicles traveling in different directions and varying environmental conditions, demanded detailed and targeted analysis to extract specific information. Despite the complexity, the Level 3 video received high scores, with 0.9 for Correctness of Information and 1.0 for Temporal Understanding, reflecting the system's ability to capture the dynamic nature and temporal changes accurately.

The most complex scenarios, represented by Level 4 and Level 5 videos, required the system to navigate even more layers of detail, including various types of vehicles, pedestrians, traffic signs, and environmental changes. These videos tested the system's capacity to maintain consistency and contextual accuracy amid a high degree of complexity. The Level 4 video, with its multiple intersecting lanes and varying traffic patterns, received a score of 0.7 for Correctness of Information and 0.6 for Detailed Orientation, indicating challenges in capturing all elements accurately. Similarly, the Level 5 video, characterized by its congested urban setting with numerous vehicles and pedestrians, scored lower with 0.6 for Correctness of Information and 0.5 for Detailed Orientation, reflecting difficulties in handling dense and complex scenes.

\begin{figure}
    \centering
    \includegraphics[width=\linewidth]{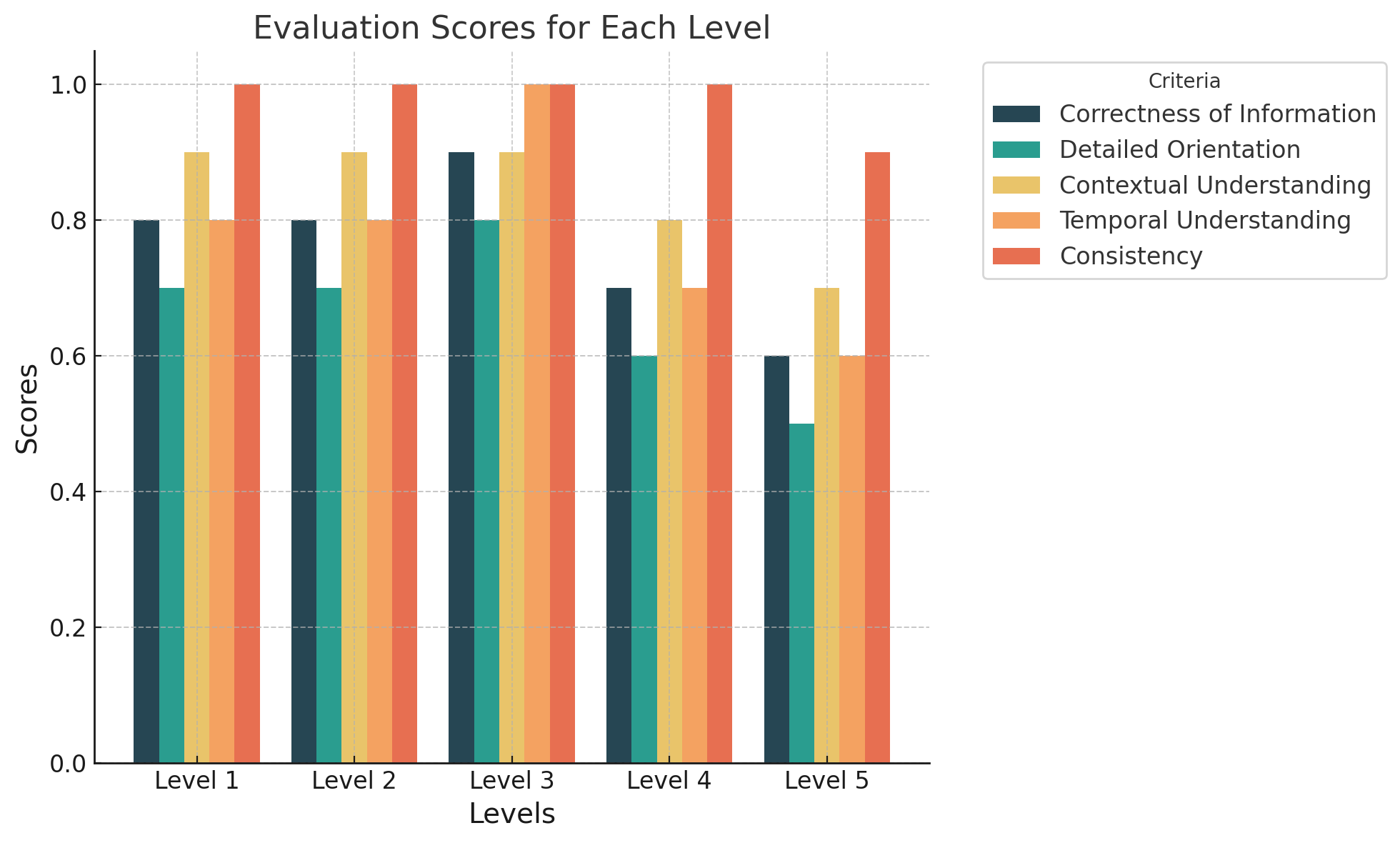}
    \caption{Summary of Evaluation Scores for each complexity level according to the criterion}
    \label{fig:eval_scores_graph}
\end{figure}

We can observe in Figure \ref{fig:eval_scores_graph}, a visible trend in the scores is the decrease as the complexity of the video content increases. This is evident in the progression from Level 1 through Level 5. For instance, the scores for Correctness of Information dropped from 0.8 in Level 1 and Level 2 to 0.6 in Level 5, reflecting the increasing difficulty in accurately capturing all relevant details in more complex scenes. Similarly, Detailed Orientation scores also decreased from 0.7 in the less complex levels to 0.5 in the most complex level, highlighting the system's challenges in maintaining detailed descriptions as complexity rises. Contextual Understanding and Temporal Understanding saw slight decreases, indicating that while the system could generally grasp the overall context and temporal changes, it struggled with the nuances as complexity increased. Consistency remained relatively high across all levels, with a slight dip to 0.9 in Level 5, demonstrating that while the system maintained a steady narrative, the detailed accuracy and depth of information were affected by the complexity.

Overall, the system showed strong consistency across all levels, with each receiving high scores in the Consistency criterion, particularly Level 1 through Level 4 scoring 1.0. This indicates that the system could maintain a steady level of detail and accuracy throughout the descriptions, regardless of the complexity of the video.

\subsection{Experiment 3: Measuring affect on performance with increasing video complexities and decreasing video quality}
\label{exp:video-quality}
Building upon the insights gained from the previous experiment, this section aims to further explore the system's performance by introducing an additional variable: video resolution. While the previous experiment focused solely on the impact of video content complexity on system latency and accuracy, this experiment extends the analysis to include the effects of varying video resolutions.

In this experiment, we evaluate the system's ability to handle different video complexity levels at four distinct resolutions: 1080p, 720p, 360p, and 144p. The goal is to determine how reducing video quality influences the system’s processing efficiency and accuracy, particularly when dealing with increasingly complex video content. By examining the interplay between video complexity and resolution, we aim to provide a comprehensive understanding of the system's robustness and adaptability under various real-world conditions. The following subsections focus on each video level complexity individually so as to provide a more comprehensive evaluation breakdown.

\subsubsection{Level 1 Complexity:}
\begin{figure}[ht]
    \centering
        \centering
        \includegraphics[width=\linewidth]{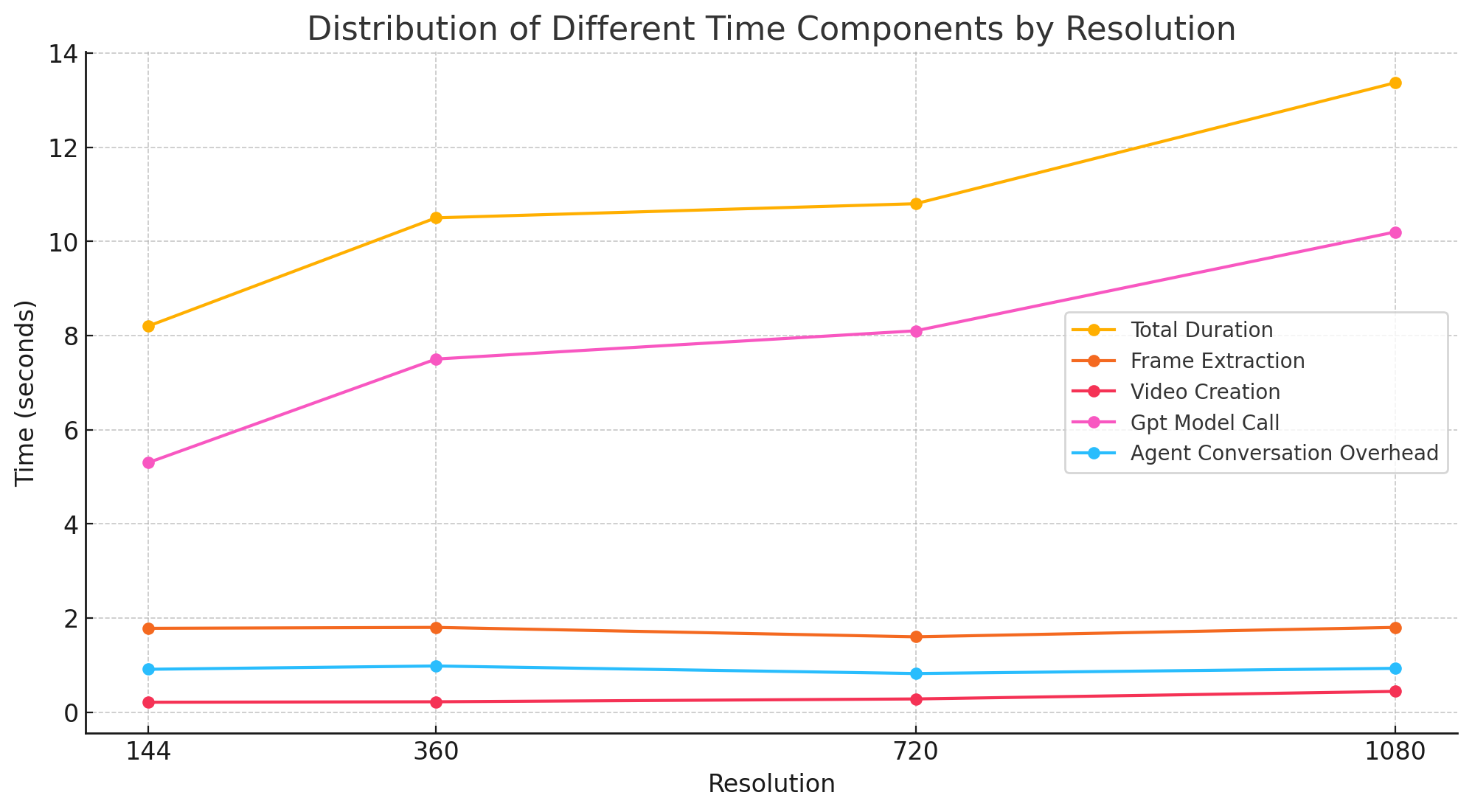}
        \caption{Evaluation Score for Level-1 Video Complexity}
        \label{level-1-eval}
\end{figure}

Figure \ref{level-1-eval} presents the latency distribution for processing a Level-1 complexity video in a two-agent system at various resolutions (1080p, 720p, 360p, and 144p). The Total Duration, representing the complete processing time, significantly decreases as the resolution lowers, from 13.37 seconds at 1080p to 8.2 seconds at 144p. This reduction is primarily driven by the decreased data volume and complexity in handling lower-resolution frames. Similarly, the GPT Model Call latency, which indicates the time taken for the GPT model to process video data and generate descriptions, shows a substantial decrease with lower resolutions: from 10.2 seconds at 1080p to 5.3 seconds at 144p. This trend highlights the increased efficiency in processing less detailed and lower-quality frames. While the times for Frame Extraction, Video Creation, and Agent Conversation Overhead exhibit minor variations, they remain relatively stable and less impactful on the overall latency. The consistent reduction in total latency and model call latency with decreasing resolution suggests that for applications prioritizing processing speed over high resolution, opting for lower resolutions can significantly enhance performance and reduce computational load..

\subsubsection{Level 2 Complexity:}
\begin{figure}[ht]
    \centering
        \centering
        \includegraphics[width=\linewidth]{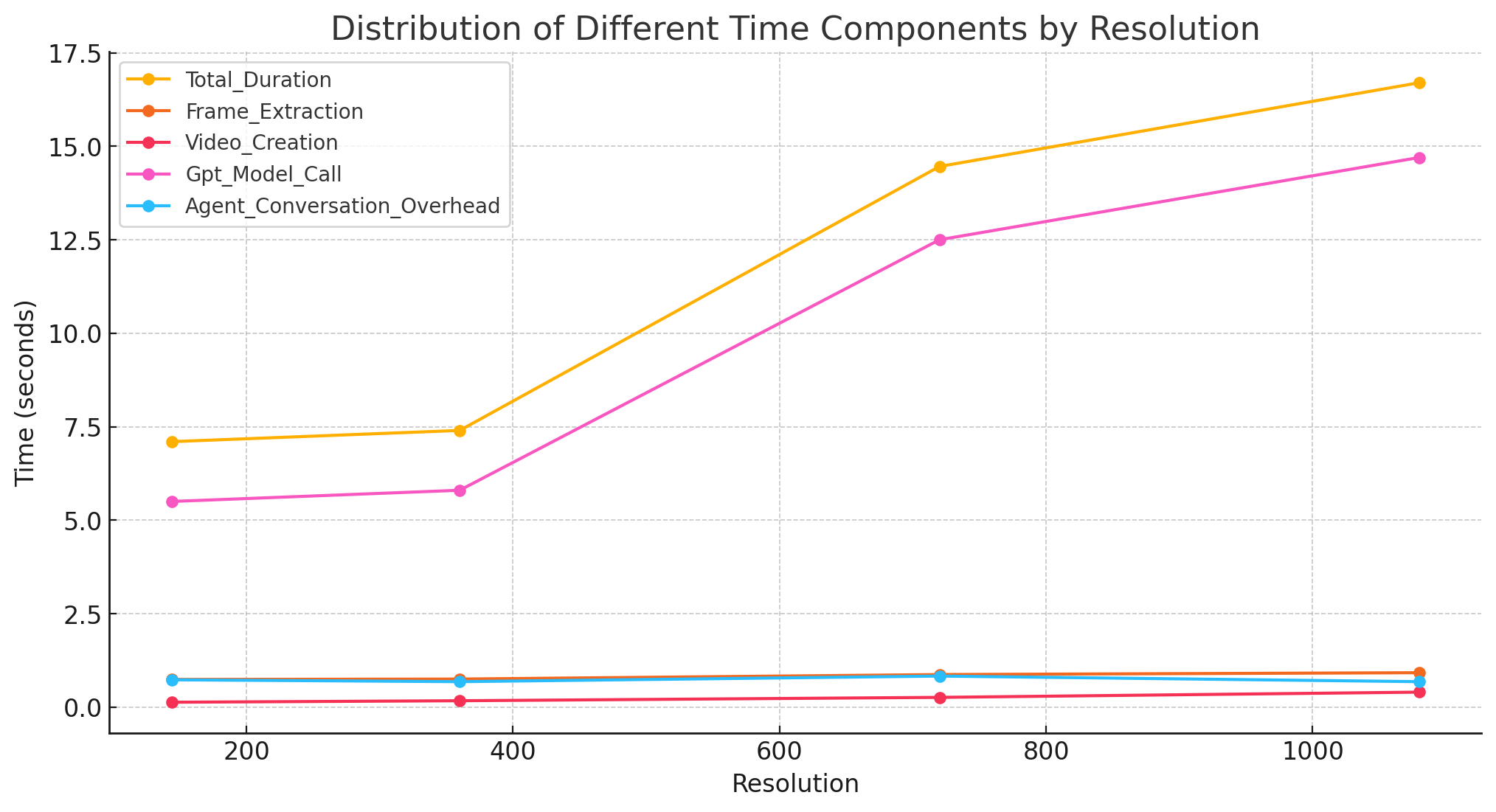}
        \caption{Evaluation Score for Level-2 Video Complexity}
        \label{level-2-eval}
\end{figure}

Figure \ref{level-2-eval} presents the latency distribution for processing a Level-2 complexity video. The Total Duration for Level 2 videos shows a significant decrease with lower resolutions, from 16.7 seconds at 1080p to 7.1 seconds at 144p. This reduction is driven by the decreased data volume at lower resolutions, making processing more efficient. The GPT Model Call latency, a critical component, also decreases from 14.7 seconds at 1080p to 5.5 seconds at 144p. This indicates that higher resolutions require more processing time due to the increased complexity and detail in the frames. Minor variations are observed in Frame Extraction, Video Creation, and Agent Conversation Overhead, but these are less impactful on the overall latency.

\subsubsection{Level 3 Complexity:}
\begin{figure}
    \centering
    \includegraphics[width=\linewidth]{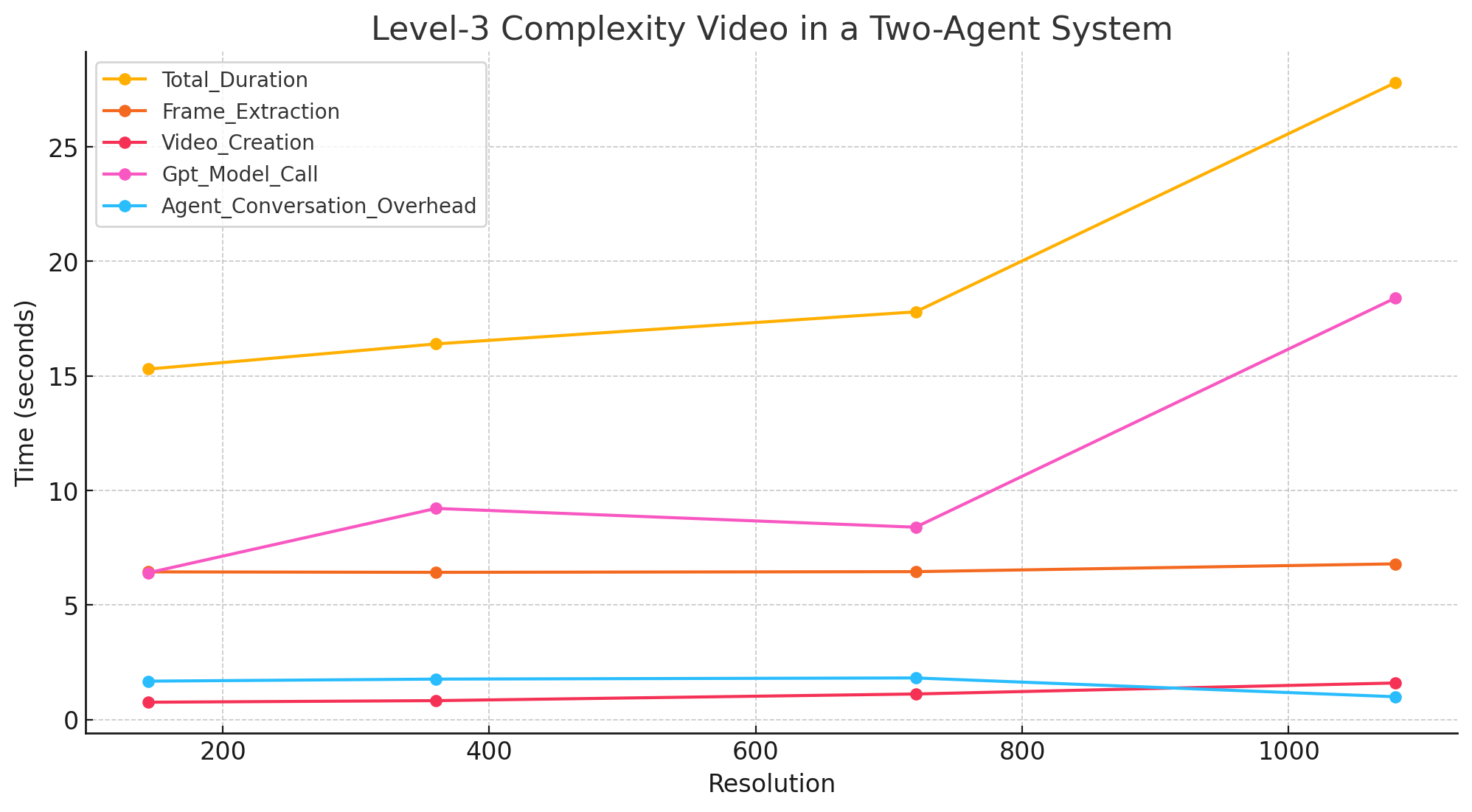}
    \caption{Processing time in the system for different resolutions of a Level-3 complexity video}
    \label{fig:level-3-eval}
\end{figure}

As can be observed in Figure \ref{fig:level-3-eval}, for Level 3 videos, the Total Duration significantly decreases from 27.8 seconds at 1080p to 15.3 seconds at 144p. The GPT Model Call latency shows a similar trend, reducing from 18.4 seconds at 1080p to 6.41 seconds at 144p. These decreases highlight the efficiency gains at lower resolutions, where less data needs to be processed. Frame Extraction, Video Creation, and Agent Conversation Overhead times exhibit minor fluctuations but remain relatively stable compared to the total latency impact. 

\subsubsection{Level 4 Complexity:}
\begin{figure}[ht]
    \centering
        \centering
        \includegraphics[width=\linewidth]{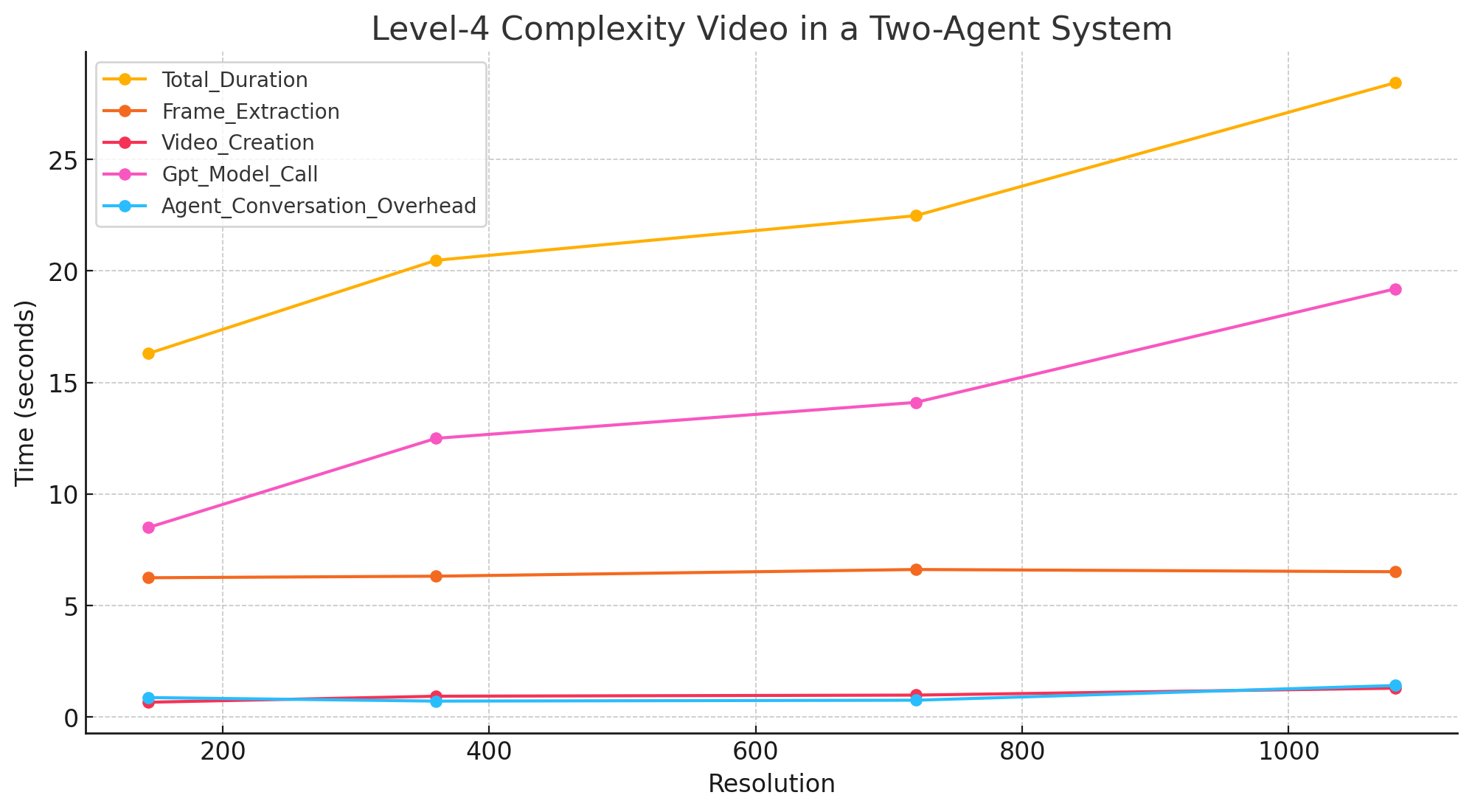}
        \caption{Evaluation Score for Level-4 Video Complexity}
        \label{level-4-eval}
\end{figure}

Figure \ref{level-4-eval} displays the Total Duration for Level 4 videos drops from 28.44 seconds at 1080p to 16.3 seconds at 144p. The GPT Model Call latency decreases significantly from 19.2 seconds at 1080p to 8.5 seconds at 144p, demonstrating the increased processing efficiency at lower resolutions. Frame Extraction and Video Creation times show minor changes, and Agent Conversation Overhead remains relatively stable, suggesting these factors have a lesser impact on the total latency compared to resolution changes.

\subsubsection{Level 5 Complexity:}
\begin{figure}[ht]
    \centering
        \centering
        \includegraphics[width=\linewidth]{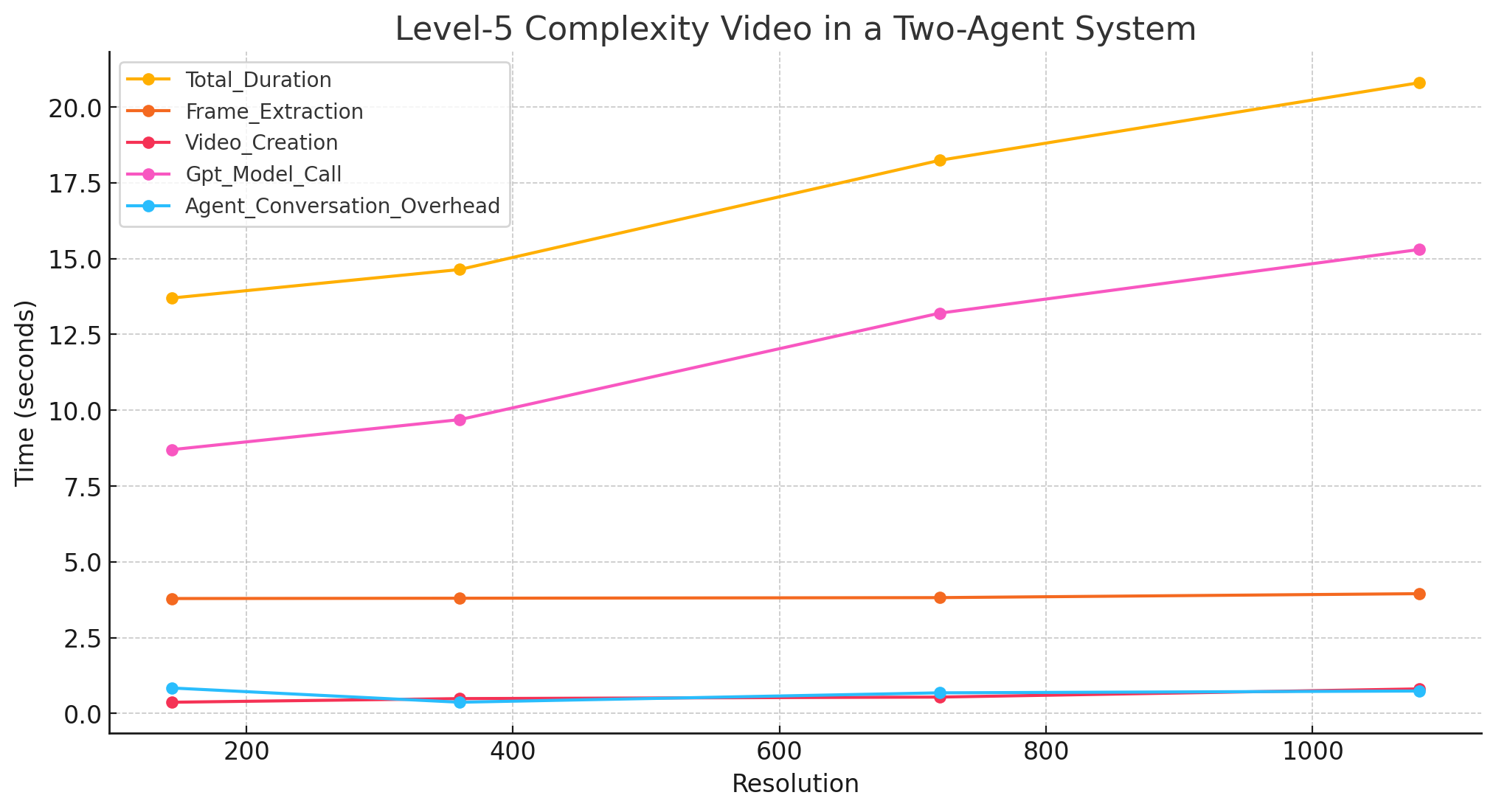}
        \caption{Evaluation Score for Level-5 Video Complexity}
        \label{level-5-eval}
\end{figure}

As can be seen Figure \ref{level-5-eval}, for Level 5 videos, the Total Duration decreases from 20.8 seconds at 1080p to 13.7 seconds at 144p. The GPT Model Call latency follows this trend, reducing from 15.3 seconds at 1080p to 8.7 seconds at 144p. These results emphasize the advantage of lower resolutions in reducing processing time. While Frame Extraction, Video Creation, and Agent Conversation Overhead times vary slightly, their impact on the overall latency is minimal compared to the significant reduction achieved by lowering the resolution.

\subsubsection{Result Summary}
Across all levels of video complexity, a clear trend is observed where lower resolutions lead to significant reductions in total latency and GPT model call latency. Higher resolutions demand more processing time due to the increased data complexity, whereas lower resolutions facilitate quicker processing by reducing the amount of data the system needs to handle. This consistent pattern highlights the importance of resolution in determining processing efficiency, especially for applications where minimizing latency is crucial. Minor variations in Frame Extraction, Video Creation, and Agent Conversation Overhead times suggest that these factors are less influential compared to the overall impact of video resolution on processing latency.

\subsection{Discussions}
The experiments highlight several key insights into the performance and applicability of the proposed LLM-based multi-agent system framework for complex event processing. The first experiment focuses on the system's ability to handle increasing numbers of agents, revealing that while adding more agents can enhance functionality and improve workflow management, it also introduces higher latency due to increased communication overhead. This trade-off between the number of agents and system latency is a critical consideration for designing scalable and efficient multi-agent systems. The optimal configuration for our specific use case was identified as four agents, balancing functionality and performance.

The second experiment evaluates the system's performance with varying levels of video complexity. It was observed that higher complexity levels resulted in decreased accuracy and increased latency, primarily due to the system's struggle to maintain detailed and accurate descriptions in more intricate scenarios. Nevertheless, the system demonstrated strong consistency across all complexity levels, indicating its robustness in maintaining a coherent narrative despite the challenges posed by complex video content.

The third experiment examines the impact of video resolution on system performance. Lower resolutions significantly reduced latency and computational load, albeit at the expense of detailed information and accuracy. This finding suggests that for applications prioritizing processing speed over high resolution, opting for lower resolutions can enhance performance and reduce the system's computational requirements.

Overall, the experiments underscore the potential of integrating LLMs with multi-agent systems to create a more flexible and robust CEP framework. The system's ability to handle diverse workflows, coupled with its adaptability to varying complexities and resolutions, demonstrates its applicability in real-world scenarios. Furthermore, the system demonstrates its usability and integration with existing tools and technologies, signifying the potential easy adoption of LLM-MASs in the technological landscape.

\subsection{Limitations}
One drawback observed from increasing the number of agents was the loss or unintended change in already present information. For example, in a two-agent system, the engineer functions as an omnipotent entity with access to all the tools and directly obtains relevant information from the user's prompt. However, with the introduction of a third agent, the system must now extract relevant information from dynamic prompts generated by the new agent based on the conversation occurring between it and the user.

\begin{figure}
    \centering
    \includegraphics[width=\linewidth]{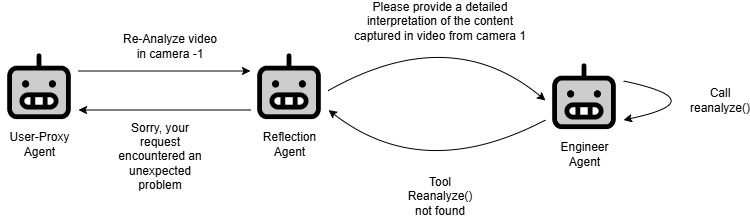}
    \caption{A three-agent system encountering an error due to an agent calling a tool that does not exist}
    \label{fig:tool-not-found}
\end{figure}

Figure \ref{fig:tool-not-found} illustrates an example scenario involving the interaction of Autogen agents to handle a video re-analysis request.The user initiates a request through the User-Proxy Agent to re-analyze video from camera 1. The Reflection Agent processes this request and delegates it to the Engineer Agent, tasked with analyzing video data. The Engineer Agent attempts to use the reanalyze() function but encounters an issue: the function does not exist. Consequently, the Engineer Agent generates an error message and sends it back to the Reflection Agent.

Another limitation of the system can be observed from the speaker selection optimization (see Section~\ref{sec:data-flow}). The speaker selection can be interpreted as a finite state machine, where each agent represents a state in the system. In the presented text, the speaker selection is optimized based on specific keywords that are added to the user prompt so that the system knows which agent must be chosen as the next speaker. This approach is manageable in a workflow involving four agents, but if the workflow is expanded to include more agents, it becomes increasingly difficult to manage. In a complex, dynamic workflow with a high number of agents, coding all possible states depending on keywords becomes a challenging task. Hence, there is a practical limit to the complexity of workflows that the system can manage.

Additionally, the current implementation does not fully leverage the potential of federated learning and edge computing, which could further enhance the system's scalability and performance. The system's dependence on predefined workflows and tool configurations may also limit its adaptability to new and unforeseen scenarios. Complex LLM-MASs is a young, and still evolving area of research. Developing more dynamic and adaptive frameworks that can learn and evolve over time would significantly enhance the system's versatility and applicability in various domains.  Future research should explore the integration of these technologies to create a more distributed and efficient CEP framework. 

\section{Conclusion}\label{sec:conclusion}
This paper explores the development and evaluation of an LLM-based multi-agent system framework for complex event processing events through a proof-of-concept that integrates state-of-the-art LLM orchestration with the publish/subscribe paradigm to address the limitations of current CEP systems. The implementation of the POC demonstrated the feasibility of using the Autogen framework in conjunction with Kafka message brokers to create an autonomous CEP pipeline. The key contributions of this work lie in the development and thorough evaluation of a proof-of-concept for an llm agent augmented publish/subscribe (pub/sub) system that seamlessly integrates large language model orchestration frameworks with established pub/sub tools like Kafka. This integration forms a novel approach to complex event processing and video analysis within distributed AI systems.


Building on the work presented above, several promising avenues for future research are identified. Enhancing inter-agent communication efficiency is a primary focus, aiming to reduce latency and improve overall system performance without sacrificing functionality. Scalability presents another critical area, where extending the framework to support a larger number of agents and more intricate workflows could be explored, potentially through hierarchical or distributed configurations. Furthermore, integrating the pub/sub system and agent tool call functionality with emerging technologies such as federated learning and edge computing could significantly expand its capabilities and practical applicability, fostering advancements in real-time data processing and decision-making within distributed AI ecosystems.

\section*{Acknowledgements}
This research is supported by the Research Council of Finland (former Academy of Finland) 6G Flagship Program (Grant Number: 346208), and by Business Finland through the Neural pub/sub research project (diary number 8754/31/2022).

\section*{Declaration of interests}
The authors declare that they have no known competing financial interests or personal relationships that could have appeared to influence the work reported in this paper.

%


\bibliographystyle{elsarticle-num}
\bibliography{preprint}

\section*{Appendix: Conversation Flow in a two-agent system in the Autogen framework}
Here we present a detailed example of a two-agent system where the agents collaborate to resolve a user query, as illustrated in Figure~\ref{fig:comp-data-flow}. The interaction between these two agents follows a consistent conversational pattern, which can be extended to scenarios involving multiple agents. For the purpose of clarity and simplicity, we focus on a two-agent conversation in this example. The structured dialogue below highlights the step-by-step communication and problem-solving process undertaken by the agents to address the user's request:
\newline

    \noindent\textbf{Admin (to chat\_manager):} \\
    What is happening in the video in camera-1?
    
    \noindent\texttt{>>>>>>>> USING AUTO REPLY...}
    
    \noindent\textbf{Engineer (to chat\_manager):} \\
    \texttt{Suggested tool call (call\_DuHMSLubhNAzz0pNbJX8QDKU): kafka\_consume } \\
    Arguments: \\
    \{"topic":"camera-1"\}
    
    \noindent\texttt{>>>>>>>> USING AUTO REPLY...}
    
    \noindent\texttt{>>>>>>>> EXECUTING ASYNC FUNCTION kafka\_consume...} \\
    
    \noindent\textbf{Admin (to chat\_manager):} \\
    \texttt{Response from calling tool (call\_DuHMSLubhNAzz0pNbJX8QDKU)} \\
    demo\_videos/Complex\_Video.mp4 \\
    
    \noindent\texttt{>>>>>>>> USING AUTO REPLY...}
    
    \noindent\textbf{Engineer (to chat\_manager):} \\
    \texttt{Suggested tool call (call\_jHJpLPJf24rgg3KFU9S3KhVD): frame\_extraction} \\
    Arguments: \\
    \{"video\_path":"demo\_videos/Complex\_Video.mp4","sampling":1,
    "output\_root":"extracted\_frames/"\}
    
    \noindent\texttt{>>>>>>>> USING AUTO REPLY...}
    
    \noindent\texttt{>>>>>>>> EXECUTING ASYNC FUNCTION frame\_extraction...} \\
    
    \noindent\textbf{Admin (to chat\_manager):}\\
    \texttt{Response from calling tool (call\_jHJpLPJf24rgg3KFU9S3KhVD)} \\
    extracted\_frames/Complex\_Video\_frames \\
    
    \noindent\texttt{>>>>>>>> USING AUTO REPLY...}
    
    \noindent\textbf{Engineer (to chat\_manager):} \\
    \texttt{Suggested tool call (call\_z3AooAxII8YiE7L8Cu0MFPNs): create\_video\_from\_frames} \\
    Arguments: \\
    \{"frames\_path":"extracted\_frames/Complex\_Video\_frames",
    "output\_video\_path":"new\_video/Complex\_Video\_result.mp4","frame\_rate":24\}
    
    \noindent\texttt{>>>>>>>> USING AUTO REPLY...}
    
    \noindent\texttt{>>>>>>>> EXECUTING ASYNC FUNCTION create\_video\_from\_frames...} \\
    Success! Video Created \ Stored in new\_video/Complex\_Video\_result.mp4
    
    \noindent\textbf{Admin (to chat\_manager):} \\
    \texttt{Response from calling tool (call\_z3AooAxII8YiE7L8Cu0MFPNs)} \\
    new\_video/Complex\_Video\_result.mp4 \\
    
    \noindent\texttt{>>>>>>>> USING AUTO REPLY...}
    
    \noindent\textbf{Engineer (to chat\_manager):} \\
    \texttt{Suggested tool call (call\_mXRaHSbTAaZaGHL9L37FMEk5): call\_model} \\
    Arguments: \\
    \{"video\_path":"new\_video/Complex\_Video\_result.mp4",
    "user\_input":"What is happening in the video in camera-1?"\}
    
    \noindent\texttt{>>>>>>>> USING AUTO REPLY...}
    
    \noindent\texttt{>>>>>>>> EXECUTING ASYNC FUNCTION call\_model...} \\

    \noindent\textbf{Admin (to chat\_manager):} \\
    \texttt{Response from calling tool (call\_mXRaHSbTAaZaGHL9L37FMEk5)} \\
    The frames from the video show a busy highway with multiple lanes of traffic. The vehicles include cars, vans, trucks, and motorcycles, all traveling in both directions. The scenes appear to be captured during dusk or early evening, as the headlights of the vehicles are on. The traffic seems to be moving smoothly without any visible congestion or accidents. The road is divided by a central barrier, and the vehicles are maintaining a steady flow.TERMINATE \\
    
    \noindent\texttt{>>>>>>>> USING AUTO REPLY...}

    \noindent\textbf{Engineer (to chat\_manager):} \\
    TERMINATE








\end{document}